\documentclass[prl,twocolumn,showpacs,floatfix,aps]{revtex4}
\usepackage{graphicx}
\newcommand{\etal}{{\it et~al.}}

\begin{document}

\title{Advances in point-contact spectroscopy: two-band
superconductor MgB$_2$ (A review) }
\author{I. K. Yanson \footnote{e-mail:
yanson@ilt.kharkov.ua}, Yu. G. Naidyuk}

\affiliation {B.Verkin Institute for Low Temperature Physics and
Engineering, National Academy  of Sciences of Ukraine, 47 Lenin
Ave., 61103, Kharkiv, Ukraine}

\begin{abstract}
Analysis of the point-contact spectroscopy (PCS) data on
the new dramatic high-T$_c$ superconductor  MgB$_2$ reveals
quite different behavior of two disconnected $\sigma$ and
$\pi$ electronic bands, deriving from their anisotropy,
different dimensionality, and electron-phonon interaction.
PCS allows direct registration of both the superconducting
gaps and electron-phonon-interaction spectral function of
the two-dimensional $\sigma$ and three-dimensional $\pi$
band, establishing correlation between the gap value and
intensity of the high-T$_c$ driving force -- the $E_{2g}$
boron vibration mode. PCS data on some nonsuperconducting
transition-metal diborides are surveyed for comparison.

\end{abstract}

\maketitle

\tableofcontents

\section{Introduction}

MgB$_2$ was discovered to be superconducting only a couple
of years ago \cite{Nagamatsu}, and despite that many of
its characteristics have now been investigated and a
consensus exists about its outstanding properties. First
of all, this refers to its high T$_c$ ($\approx $ 40\,K)
which is a record-breaking value among the {\it s-p}
metals and alloys. It appears that this material is a rare
example of multi-band (at least two) electronic structure,
which are weakly connected with each other. These bands
lead to very uncommon properties. For example, T$_c$ is
almost independent of elastic scattering, unlike for other
two-band superconductors \cite{Mazin}. The maximal upper
critical magnetic field can be made much higher than that
for a one-band dirty superconductor \cite{Gurevich}. The
properties of MgB$_2$ have been comprehensively calculated
by the modern theoretical methods, which lead to a basic
understanding of their behavior in various experiments.

\subsection{Crystal structure}

\begin{figure}
\includegraphics[width=6cm,angle=0]{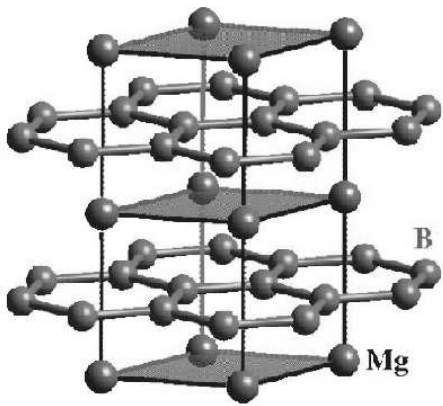}
\caption[]{Crystal structure of MgB$_2$.} \label{crystal}
\end{figure}

Magnesium diboride, like other diborides MeB$_2$ (Me=Al, Zr, Ta,
Nb, Ti, V etc.), crystalizes in a hexagonal structure, where
honeycomb layers of boron are intercalated with hexagonal layers
of magnesium located above and below the centers of boron hexagons
(Fig. \ref{crystal}).
The bonding between boron atoms is much stronger than that
between magnesium, and therefore the disordering in the
magnesium layers appears to be much easier than in the
boron layers. This difference in bonding between boron and
magnesium atoms hinders the fabrication of MgB$_2$ single
crystals of appreciable size.

\subsection{Electron band structure}

\begin{figure}
\includegraphics[width=8cm,angle=0]{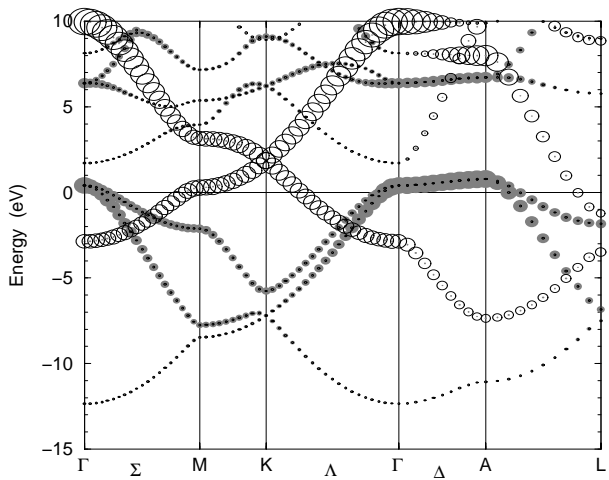}
\caption[]{Band structure of MgB$_2$ with the B $p$-character. The
radii of the hollow (filled) circles are proportional to the $\pi$
($\sigma$) character and zero-line marks the Fermi energy. After
Mazin \etal \cite{Mazin1}.} \label{eldisp}
\end{figure}

The electron band structure of MgB$_2$ has been calculated
using different {\it ab initio } methods yielding basically
the same result \cite {An,Kong,Kortus,Liu,Yildirim}. The
$E({\rm k})$ curves  are shown in Fig.\,\ref{eldisp}.

The dispersion relations are shown for boron $p$-character
orbitals, which play a major role in transport and thermodynamic
properties. The radii of the hollow circles are proportional to
the $\pi $-band character, which is made from $p_z$ boron
orbitals,
while those of the filled circles are proportional to the $%
\sigma $-band character, made from $p_{xy}$ orbitals. The most
important is a quasi two-dimensional dispersion relation along the
$\Gamma $A $(\Delta )$ direction with a small Fermi energy
$\approx $0.6 eV, and accordingly, with a moderate Fermi velocity.
The corresponding sheets of the Fermi energy form the cylindrical
surfaces along the $\Gamma $A direction seen in Fig.\,\ref{Choi-D}
below. The corresponding electron transport is very anisotropic
($\rho_c/\rho_{ab}\simeq 3.5$ \cite{Eltsev}) with the plasma
frequency for the $\sigma$ band along the $c$ (or $z$) axis being
much smaller than that in the $ab$ $(xy)$ direction
\cite{Brinkman}. The hole branch along $\Gamma $A experiences a
huge interaction with the phonon $E_{2g}$ mode for carriers moving
along the $ab$ plane (see below), although its manifestation is
screened effectively by the much faster hole mobility in the $\pi
$-band \cite{Mazin}.

In a dirty material, with prevailing disorder in the magnesium
planes, the $\pi $-band conductivity is blocked by defects, and
the $\sigma $ band takes over, implying greater electron phonon
interaction (EPI) than in the clean material. This constitutes a
plausible explanation for the violation of the Matthiesen rule,
which manifests itself in an increase of  the residual resistivity
together with an increase of the temperature coefficient at a high
temperature \cite{Mazin}.

At the same time, the critical temperature $T_c$ does not
decrease substantially in dirty materials  \cite{Mazin},
since the superconductivity is induced by EPI in the
$\sigma $-band, whose crystal order is much more robust.

This consideration is very important in understanding the
point-contact data, since the disorder at the surface of
the native sample depends on the position of the contact
spot, and because of the uncontrolled introduction of the
further disorder while fabricating the contact.

\subsection{Critical magnetic field}

In a clean material the layered crystal structure dictates
strong anisotropy of the upper critical magnetic fields
$B_{c2}^{ab}\gg B_{c2}^{c}$. Their ratio at low
temperatures reaches about 6 while $B_{c2}^{c}$ is as low
as 2--3 T \cite{Kogan}. If the field aligned is not
precisely along the $ab$ plane, the $B_{c2}$ value is
strongly decreased.

On the other hand, for a dirty material the anisotropy is
decreased (to a ratio of about 1.6$\div $2), but both the
magnitudes of $B_{c2}^{ab}$ and $B_{c2}^{c}$ are strongly
increased. For strongly disordered sample, it may be as high as
40\,T \cite{Gurevich}! It is interesting that this high value is
achieved at low temperature, where the disordered $\pi $ band is
fully superconducting.

Hence, we may expect that the value of critical magnetic
field at low temperatures is the smaller the cleaner is the
part of the MgB$_2$ volume near the contact, provided its
$T_c\simeq T_c^{bulk}$. This observation is important in
the classification of contacts with respect to their
purity.

\subsection{Phonons and Electron-Phonon Interaction}

\begin{figure}
\includegraphics[width=8cm,angle=0]{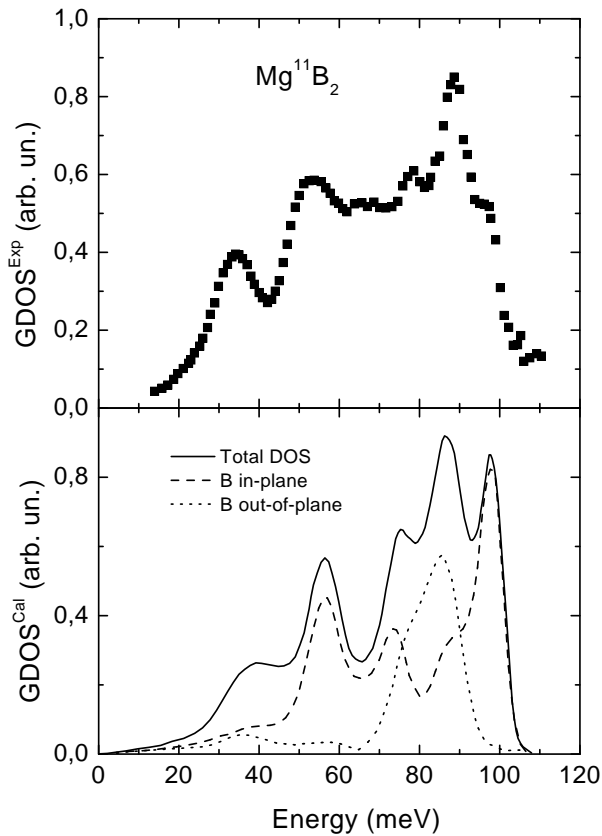}
\caption[]{Upper panel: Phonon density of states in MgB$_2$
determined experimentally  by neutron scattering. Bottom
panel: calculated curve (solid line) with decomposition on
boron atoms vibrating out of $ab$ plane (dotted curve) and
parallel to it (dashed curve). After Osborn \etal
\cite{Osborn}.} \label{phononDOS}
\end{figure}

\begin{figure}
\includegraphics[width=8cm,angle=0]{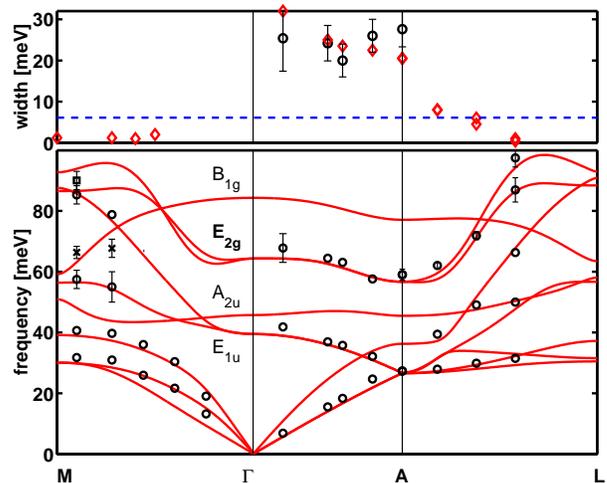}
\caption[]{Dispersion curves of phonons  in MgB$_2$ and the width
of phonon lines determined by inelastic X-ray scattering (symbols)
together with calculations (solid lines). After Shukla \etal
\cite{Shukla}.} \label{Shukla1}
\end{figure}

The phonon density of states (PDOS) is depicted in Fig.
\ref{phononDOS}. The upper panel shows the measured PDOS at
$T=8$\,K, while the lower ones shows the calculated DOS with the
partial contribution from boron atoms moving in the $ab$-plane and
out of it. One can see the peak for boron atoms moving in the $ab$
plane at $\simeq 75$ meV, which plays a very important role in the
electron-phonon interaction, as is shown in Fig.\,\ref{Shukla1},
measured by inelastic $X$-ray scattering \cite{Shukla}. This mode
gives a weakly dispersion branch between 60 and 70 meV in the
$\Gamma $A direction with $E_{2g}$ symmetry at the $\Gamma $
point. The linewidth of this mode is about 20$\div $28 meV along
the $\Gamma $A direction, while along the $\Gamma $M direction it
is below the experimental resolution. The same phonon peak is
active in Raman scattering \cite{Bohnen,Quilty,Goncharov}. It is
located at the same energy with the same linewidth. This points to
the very strong EPI for this particular lattice vibration mode.
The same result follows from theoretical considerations.

\begin{figure}
\includegraphics[width=7cm,angle=0]{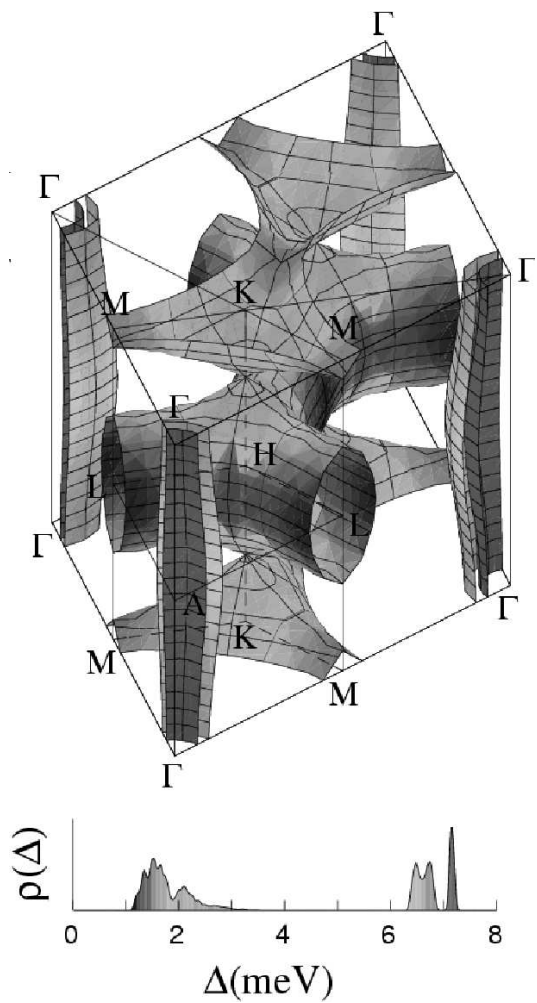}
\caption[]{Superconducting energy gap distribution over
the Fermi surface (FS) of MgB$_2$. The gap value around 7
meV corresponds to cylinder like sheets of the FS centered
at $\Gamma$ points, while the small gap value around 2 meV
corresponds to the tubular FS network. After Choi \etal
\cite{Choi}. } \label{Choi-D}
\end{figure}

Figure \ref{Choi-D} shows the distribution of the
superconducting energy gap on the Fermi surface of MgB$_2$
\cite{Choi}. The maximum gap value is calculated along the
$\Gamma $A direction due to the very strong EPI. Just in
this direction is located 2D $\sigma $ band (cylinders
along the $\Gamma $A direction). The 3D $\pi $ band has a
much smaller EPI, and, correspondingly, a smaller energy
gap. The EPI parameter $\lambda $ can be decomposed between
different pieces of the Fermi surface.
It is shown \cite{ChoiPRB} that the value of $\lambda$ on
the $\sigma $ band amounts to $2\div 3$. Moreover,
$\lambda_\sigma $ can be decomposed between different
phonon modes, and it appears that only the E$_{2g}$ phonon
mode along the $\Gamma $A direction plays a major role with
a partial $\lambda_{\sigma} $ value of about $\simeq 25$
\cite{An1}, though concentrated in a very restricted phase
space.

\subsection{Mechanism for high $T_c$ in MgB$_2$}

The commonly accepted mechanism for high $T_c$ in MgB$_2$
is connected with the strong interaction between charge
carriers and phonons in the $E_{2g}$ mode. This mode is due
to antiparallel vibration of atoms in the boron planes.
The key issue is that along the $\Gamma $A direction the
electron band structure is such that the Fermi energy of
the hole carriers is only 0.5$\div 0.6$\,eV, which shrinks
even more when the borons deviate from the equillibrium
positions. Together with the 2D structure of the
corresponding sheet of the Fermi surface, this leads to a
constant density of states at the Fermi energy and,
correspondingly, to very large EPI with partial $\lambda
_\sigma $ (the EPI parameter in the $\sigma $ band) of
about $\sim 25$ \cite{An1}. Cappelluti {\it et al.
}\cite{Cappelluti} point out that the small Fermi velocity
for charge carriers along the $\Gamma $A direction leads
to a large nonadiabatic correction to $T_c$ (about twice
as much compared with the adiabatic Migdal-Eliashberg
treatment). Although this interaction is a driving force to
high $T_c$ in this compound, it does not lead to crystal
structure instability, since it occupies only a small
volume in the phase space.

The role of the $\pi $ band is not completely clear. On the one
hand, the $\pi $ and $ \sigma $ bands are very weakly connected,
and for some crude models they can be thought as being completely
disconnected. On the another hand, the energy gap of the $\pi $
band goes to zero at the same $T_c$ as in the bulk, and
correspondingly $2\Delta _\pi (0)/kT_c=1.4$, which is much less
than the value predicted by the weak coupling BCS theory. One can
think of the $\pi $ band as having intrinsically much lower
$T_c\approx 10$ K than the bulk \cite{Bouquet}, and at higher
temperatures its superconductivity is induced by a proximity
effect in {\bf k}-space from $\sigma $ band \cite{YansonPRB}. This
proximity effect is very peculiar. On the one hand, this proximity
is induced by the interband scattering between the $\pi $ and
$\sigma $ sheets of the Fermi surface. On the other hand, the
charge carriers connected with the $\pi $ band are mainly located
along the magnesium planes, which can be considered as a proximity
effect in coordinate space for alternating layers of $S-N-S$
structure, although on a microscopic scale. Moreover, many of the
unusual properties of MgB$_2$ may be modelled by an alternating
$S-N-S$ layer structure, the limiting case to the crystal
structure of MgB$_2$. In other words, MgB$_2$ presents a crossover
between two-band superconductivity and a simple proximity effect
structure.

\section{Samples}

We have two kind of samples supplied for us by our
colleagues from the Far-East. \footnote{The films were
provided by S.-I. Lee from National Creative Research
Initiative Center for Superconductivity, Department of
Physics, Pohang University of Science and Technology,
Pohang, South Korea. The single crystals were provided by
S. Lee from Superconductivity Research Laboratory, ISTEC,
Tokyo, Japan.}
\begin{figure}
\includegraphics[width=8cm,angle=0]{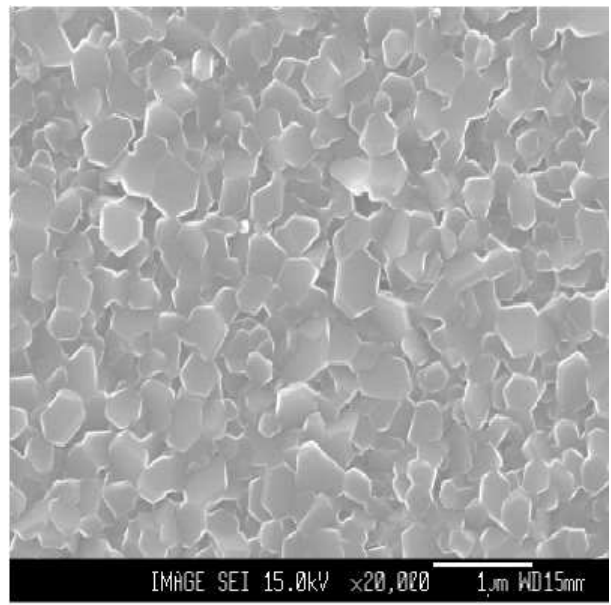}
\caption[]{Scanning electron microscopy image of MgB$_2$ films.
After Kang \etal \cite{Sung-Ik}.} \label{films}
\end{figure}

The first is a thin film with a thickness of about several
hundred of nanometers (Fig.\,\ref{films}) \cite{Kang}.
Similar films have been investigated by several other
groups with different methods. These films are oriented
with their {\it c}-axis  perpendicular to the substrate.
The residual resistance is about several tens of $\mu
\Omega \,$cm with a residual resistance ratio (RRR) $
\simeq 2.2$. This means that on average the films have a
disorder between crystallites.

 It does not exclude the possibility that on some spots the films contain clean
enough small single crystals on which we occasionally may
fabricate a point contact: see Fig. \ref{films}. Normally,
we make a contact by touching the film surface by noble
metal counter electrode (Cu, Au, Ag) in the direction
perpendicular to the substrate. Thus, nominally the
preferential current direction in the point contact is
along the {\it c} axis. Nevertheless, since the surface of
the films contains terraces with small crystallites, point
contact to the {\it ab} plane of these crystallites is also
possible. Sometimes, in order to increase the probability
of making the contact along the {\it ab} plane, we broke
the substrate with the film and made contact to the side
face of the sample.

\begin{figure}
\includegraphics[width=7cm,angle=0]{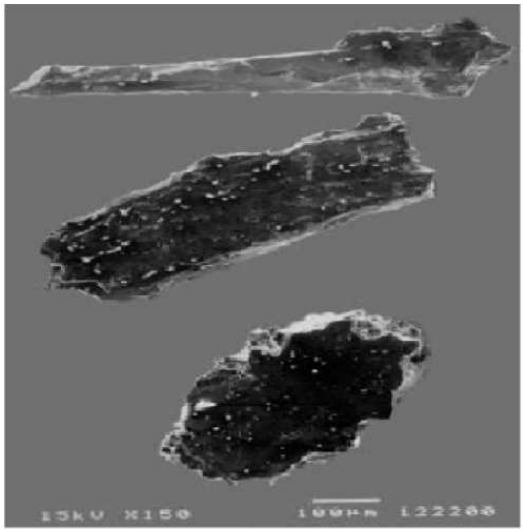}
\caption[]{Scanning electron microscopy image of MgB$_2$ single
crystals. After Lee \etal \cite{Lee}.} \label{scrys}
\end{figure}

The second type of sample is single crystal \cite{Lee}
which also was measured by other groups
\cite{Eltsev,Quilty1}. Crystals are plate-like (flakes)
and have sub-millimeter size (see Fig.\,\ref{scrys}). They
were glued by silver epoxy to the sample holder by one of
their side faces. The noble metal counter electrode was
gently touched in liquid helium by another (the opposite)
side face of the crystal. In this way we try to
preferentially make a contact along the {\it ab} plane. On
average, in the bulk, the single crystals are cleaner than
the films, but one should be cautious, since the
properties of the crystal surface differ from the
properties of the bulk, and fabrication of a point contact
may introduce further uncontrolled  defects into the
contact area.

Thus, {\it a priori} one cannot define the structure and
composition of the  contacts obtained. Nevertheless, much of that
information can be ascertained by measuring various
characteristics of a contact. Among those the most important is
the Andreev-reflection-non-linearities of the $I-V$ curves in the
superconducting energy-gap range. The magnetic field and
temperature dependences of the superconducting non-linearities
supply us with additional information. And finally, much can be
extracted from the $I-V$ nonlinearities in the normal states (the
so-called point-contact spectra). The more information we can
collect about the electrical conductivity for different conditions
of the particular contact, the more detailed and defined picture
of it emerges. It is not an easy task, since a contact has limited
lifetime, due to electrical and mechanical shocks.

Let us make a rough estimate of the distance scales involved in
the problem. The crystallite size of films is of the order of
100\,nm (see \cite{Sung-Ik}). The contact size $d$ in the
ballistic regime equals $d\simeq\sqrt{\rho l/R}$ (the Sharvin
formula). Taking $\rho l\cong 7\,10^{-7}\Omega \,$cm$\times
7\,10^{-6}$cm$=5\times 10^{-12} \Omega \,$cm$^2$ \cite{Eltsev}, we
obtain $d\simeq 7$ nm both along the $ab$ and $c$ directions for
typical resistance of 10 $\Omega$. If we suppose that a grain is
dirty (with very short mean free path), then we apply the Maxwell
formula $d\sim\rho/R$ with the results for $d$ about 0.7\,nm and
2.6\,nm for $ab$ and $c$ directions, respectively, taking $\rho$
for corresponding directions from the same reference
\cite{Eltsev}. Thus, the contact size can be of the order or
smaller than the electronic mean free path ($l_{ab}=70$ nm and
$l_c=18$\,nm, according to \cite{Eltsev}), which means that we are
working, apparently, in the spectroscopic regime, probing only a
single grain.

Rowell \cite{Rowell}, analyzing a large amount of
experimental data for the resistivity and its temperature
dependence, came to the conclusion that for highly
resistive samples only a small part of the effective cross
section should be taken into account. The reason is that
the grains in MgB$_2$ are to great extent disconnected by
oxides of magnesium and boron. For point-contact
spectroscopy previous analysis leads us to the conclusion
that the contact resistance is frequently measured only
for a single grain or for several grains, with their
intergrain boundaries facing the contact interface. This
is due to the current spreading on a scale of the order of
the contact size $d$ near the constriction.

\section{Theoretical background of PCS}

\subsection{Non-linearity of $I-V$ characteristic}

The non-linearities of the $I-V$ characteristic of a
metallic contact, when one of the electrodes is in the
superconducting state, can be written as \cite {Omel,Khlus}

\begin{equation}
I\left( V\right) \simeq \frac V{R_0}-\delta I_{ph}^N(V)+I_{exc}(V)
\label{I-V}
\end{equation}
Here $R_0$ is the contact resistance at zero bias in the
normal state. $\delta I_{ph}^N(V)$ is the backscattering
inelastic current which depends on the electron mean free
path (mfp) $l$. For ballistic contact this term is equal,
in order of magnitude, to

\begin{equation}
\delta I_{ph}^N(V)\sim \frac d{l_{in}}I(V)  \label{inelastic}
\end{equation}
where $l_{in}$ is the inelastic electron mfp, and $d$ is
the  characteristic contact
diameter. If the electron flow through the contact is diffusive ($%
l_{el}\ll d$, $l_{el}$ being an elastic mfp) but still spectroscopic, since $%
\sqrt{l_{in}l_{el}}\gg d$, then the expression
(\ref{inelastic}) should be multiplied by $l_{el}/d$. This
decreases the characteristic size, for which the inelastic
scattering is important, from $d$ to $l_{el}$
($d\rightarrow l_{el}$), and for short $l_{el}$ makes the
inelastic current very small. We notice that the inelastic
backscattering current $\delta I_{ph}^N(V)$ in the
superconducting state is approximately equal to the same
term in the normal state. Its second derivative turns out
to be directly proportional to the EPI function
$\alpha^2(\omega)\,F(\omega)$ \cite{KOS,YansonSC}
\begin{equation}
\label{pcs} -\frac{d^2I}{dV^2}\propto \frac{8\,ed}{3\,\hbar v_{\rm
F}}\alpha^2(\omega)\,F(\omega)
\end{equation}
where $\alpha$ describes the strength of the electron
interaction with one or another phonon branch, and
$F(\omega)$ stands for the phonon density of states. In
point-contact (PC) spectra the EPI spectral function
$\alpha^2(\omega)\,F(\omega)$ is modified by the transport
factor, which strongly increases  the backscattering
processes contribution.

In the superconducting state the excess current
$I_{exc}$\,(\ref{I-V}), which is due to the Andreev
reflection of electron quasiparticles from the N-S
boundary in a N-c-S point contact (c stands for
''constriction''), can be written as

\begin{equation}
I_{exc}\left( V\right) =I_{exc}^0+\delta I_{exc}\left( V\right)
\label{excess}
\end{equation}
where $I_{exc}^0\approx \Delta /R_0\approx$ const for
$eV>\Delta $ ($\Delta $ being the superconducting energy
gap).

The nonlinear term in the excess current (\ref{excess})
can be decomposed in its turn in two parts, which depend
in different ways on the elastic scattering of electron
quasiparticles:

\begin{equation}
\delta I_{exc}\left( V\right) =\delta I_{exc}^{el}\left( V\right)
+\delta I_{exc}^{in}\left( V\right)  \label{excess1}
\end{equation}
where $\delta I_{exc}^{el}\left( V\right) $ is of the
order of $\left( \Delta /eV\right) I_{exc}^0$, and $\delta
I_{exc}^{in}\left( V\right) \sim \left( d/l_{in}\right)
I_{exc}^0$. Notice that the latter behaves very similar to
the inelastic backscattering current $\delta I_{ph}^N(V)$,
namely, it disappears if $l_{el}\rightarrow 0$, while the
first term in the right hand side of expression
(\ref{excess1}) does not depend on $l_{el}$ in the first
approximation. This enables one to distinguish the elastic
term from the inelastic. Finally, all excess current terms
disappear when the superconductivity is destroyed, while
$\delta I_{ph}^N(V)$ remains very similar in both the
superconducting and normal states.

The expression for the elastic term in the excess current was
calculated for {\it ballistic} N-c-S contacts by Omelyanchuk,
Kulik and Beloborod'ko \cite {Om-Kul-Bel}. Its first derivative
equals $(T=0)$:

\begin{equation}
\left( \frac{dI_{exc}^{el}}{dV}\right) _{NcS}^{ballistic}=\frac 1{R_0}\left| \frac{%
\Delta (eV)}{eV+\sqrt{\left( eV\right) ^2-\Delta ^2\left( eV\right) }}%
\right| ^2  \label{Om-Kul-Bel}
\end{equation}

For the {\it diffusive} limit $\left( l_i\ll d\right) ,$
Beloborod'ko and Kulik derived the current-voltage
characteristic (see Eq.(21) in Ref.\cite {Bel-Om}), which
for the first derivative at $T=0$ gives \cite{Bel}:
\begin{eqnarray}
\lefteqn{R_0\left( \frac{dI_{exc}^{el}}{dV}\right) _{NcS}^{diffusive}=\nonumber}\hspace{7cm}\\
=\frac 12\ln \left| \frac{%
eV+\Delta (eV)}{eV-\Delta (eV)}\right| \frac{\Re{\left( \frac{eV}{\sqrt{%
(eV)^2-\Delta ^2(eV)}}\right) }}{\Re{\left( \frac{\Delta (eV)}{\sqrt{%
(eV)^2-\Delta ^2(eV)}}\right)} }  \label{Bel} \label{brinkeq}
\end{eqnarray}

For the sake of comparison, the similar expression of the
nonlinear term in  NIS tunnel junctions (I stands for
''insulator''), due to the self-energy superconducting
energy gap effect, has the form \cite{Wolf}:

\begin{equation}
\left( \frac{dI}{dV}\right) _{NIS}=\frac 1{R_0}\Re{\left[ \frac{eV}{%
\sqrt{\left( eV\right) ^2-\Delta ^2\left( eV\right) }}\right]}
\label{Wolf}
\end{equation}
Equations (\ref{Om-Kul-Bel}), (\ref{Bel}), and
(\ref{Wolf}) are identical in their structure and take
into account the same effect, viz., the renormalization of
the energy spectrum of a superconductor in the vicinity of
characteristic phonon energies.

From the expressions (\ref{I-V}), (\ref{inelastic}),
(\ref{excess}), and (\ref {excess1}) it becomes clear that
only on the relatively {\it clean} spots one can observe
the inelastic backscattering current $\delta I_{ph}^N(V)$
provided that the excess current term $\delta
I_{exc}^{in}\left( V\right) $ is negligible. The latter
can be canceled by suppression of superconductivity either
with magnetic field or temperature. On the contrary, in
the superconducting state, for {\it dirty} contacts, all
the inelastic terms are very small, and the main
nonlinearity is provided by the $\Delta (eV)$-dependence
of the excess current (\ref{Bel}).

\subsection{Two-band anisotropy}

Brinkman {\it et al.} have shown \cite{Brinkman} that in
the clean case for an NIS MgB$_2$ junction, the normalized
conductance is given by
\begin{equation}
\sigma \left( V\right) =\frac{\left( \frac{dI}{dV}\right)
_{NIS}}{\left( \frac{dI}{dV}\right) _{NIN}}=\frac{\left( \omega
_p^\pi \right) ^2\sigma _\pi ^{}\left( V\right) +\left( \omega
_p^\sigma \right) ^2\sigma _\sigma ^{}\left( V\right) }{\left(
\omega _p^\pi \right) ^2+\left( \omega _p^\sigma \right) ^2}
\label{Brinkman}
\end{equation}
where $\omega _p^{\pi (\sigma )}$ is the plasma frequency
for the $\pi (\sigma )$ band and $\sigma _{\pi (\sigma
)}^{}\left( V\right) $ is the normalized conductivity of
the $\pi (\sigma )$ band separately. The calculated
tunneling conductance in the {\it ab}-plane and along the
{\it c}-axis are \cite{Brinkman}
\begin{eqnarray}
\sigma _{ab}\left( V\right) &=&0.67\sigma _\pi \left( V\right) +0.33\sigma
_\sigma \left( V\right) \\
\sigma _c\left( V\right) &=&0.99\sigma _\pi \left( V\right)
+0.01\sigma _\sigma \left( V\right) \label{brinkeq2}
\end{eqnarray}

Hence, even along the {\it ab}-plane the contribution of
the $\sigma $ band is less than that of the $\pi $ band,
 to say nothing about the direction along the {\it c} axis,
where it is negligible small. The calculation predicts
that if the ''{\it tunneling cone}'' is about several
degrees from the precise {\it ab} plane, then the two
superconducting gaps should be visible in the tunneling
characteristics. In other directions only a single gap,
corresponding to the $\pi $ band, is visible. We will see
below that this prediction is fulfilled in a point-contact
experiment, as well.

Things are even worse when one tries to measure the
anisotropic Eliashberg function by means of
superconducting tunneling. The single-band numerical
inversion program \cite{Wolf,Parks} gives an uncertain
result, as was shown in Ref.\,\cite{Dolgov}.

Point-contact spectroscopy in the normal state can help in this
deadlock situation. It is known that the inelastic backscattering
current is based on the same mechanism as an ordinary homogeneous
resistance, provided that the maximum energy of the charge
carriers  is controlled by an applied voltage. The electrical
conductivity of MgB$_2$ can be considered as a parallel connection
of two channels, corresponding to the $\pi $ and $ \sigma $ bands
\cite{Mazin}. The conductivity of the $\pi $ band can be blocked
by disorder of the Mg-atoms. This situation is already obtained in
experiment, when the temperature coefficient of resistivity
increases simultaneously with an increase of the residual
resistivity, which leads to violation of Matthiessen's rule (see
Fig.\,3 in \cite{Mazin}). In this case we obtain the direct access
to the $\sigma $-band conductivity, and the measurements of the PC
spectra of the EPI for the $\sigma$ band is explicitly possible in
the normal state. Below we will see that this unique situation
happens in single crystals along {\it ab} plane.

\section{Experimental results}

\subsection{Superconducting energy gaps}

\subsubsection{{\it c}-axis oriented thin films}

\begin{figure}
\includegraphics[width=8cm,angle=0]{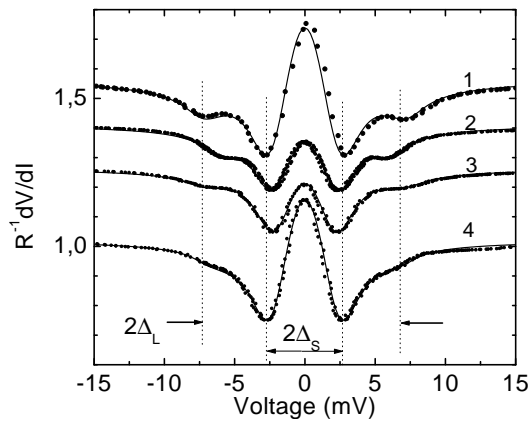}
\caption[]{Typical shapes of $dV/dI$ (experimental dots) for 4
contacts between MgB$_2$ thin film and Ag with the corresponding
BTK fitting (lines) \cite{Naidyuk1}. $\Delta_{L(S)}$ stand for
large (small) superconducting energy gap. After Naidyuk \etal
\cite{Naidyuk1}.} \label{MgB2del}
\end{figure}

\begin{figure}
\includegraphics[width=8cm,angle=0]{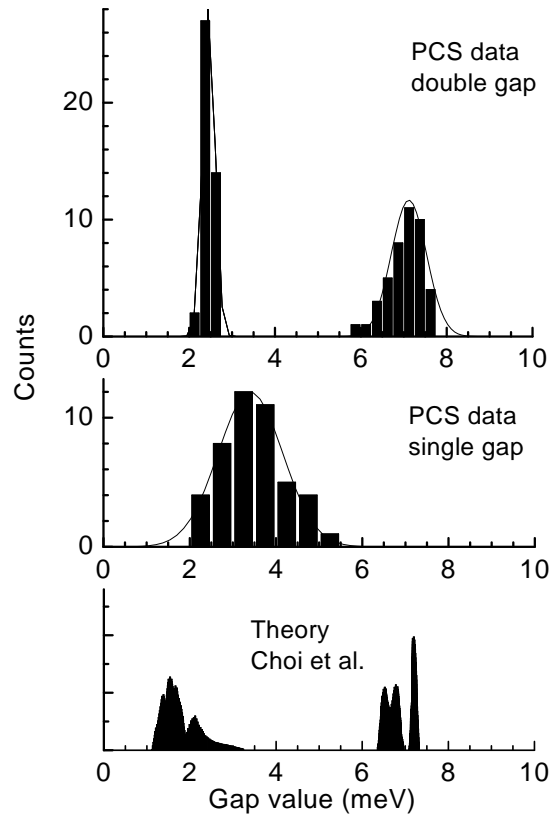}
\caption[]{Superconducting energy gap distribution of about 100
different junctions prepared on a MgB$_2$ c-axis oriented film. On
the lower panel the theoretical distribution is shown. After
Naidyuk \etal \cite{Naidyuk1}.} \label{MgB2hist}
\end{figure}

Our measurements of the superconducting energy gap by means of
Andreev reflection from about a hundred NcS junctions yield two
kinds of $dV/dI$-curve, shown in Fig.\,\ref{MgB2del}.
The first one clearly shows two sets of energy gap minima located,
as shown in distribution graph of Fig.\,\ref{MgB2hist} (upper
panel), at 2.4$\pm $ 0.1 and 7.1$\pm $0.4 meV. These curves can be
nicely fitted by BTK \cite{BTK} theory (with small $\Gamma$
parameter) for two conducting channels with an adjusted gap
weighting factor \cite{Naidyuk1}. The second kind is better fitted
with a single gap provided an increased depairing parameter
$\Gamma $ (Fig.\,\ref{MgB2hist} (middle panel)). Certainly, the
division of the gap structure into two kinds mentioned is
conventional, and depends upon the circumstance that the larger
energy gap is explicitly seen. These two kinds of gap structure
comprise about equal parts of the total number of junctions.
Usually the contribution of the large gap in the double-gap
spectra is an order of magnitude lower than that of the small one,
which is in line with the small contribution of the $\sigma$ band
into the  conductivity along the c axis (see Eq.\,\ref{brinkeq2}).

\begin{figure}
\includegraphics[width=8cm,angle=0]{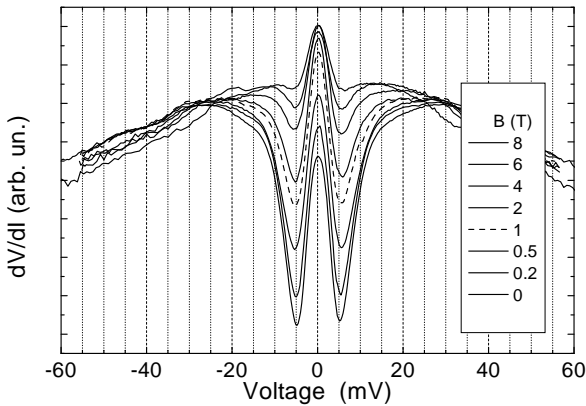}
\caption[]{Negative slope of $dV/dI$ at large biases for a 36
$\Omega$ contact between MgB$_2$ single crystal and Ag showing the
magnetic-field gap-structure evolution at 4.2\,K.}
\label{negative}
\end{figure}

It is important to note that the critical temperature of the
material around the contact is not more than a few K below $ T_c$
in the bulk material. This is determined by the extrapolating the
temperature dependence of PC spectra up to the normal state. Such
an insensitivity of $T_c$ on the elastic scattering rate is
explained in Ref. \cite{Mazin}. Nevertheless, we stress that the
gap structure (either double- or single-gap feature, and the
position of the single-gap minimum on $dV/dI$) depends very much
on random variation of the scattering in the contact region.
Moreover, since the main part of the junction conductivity is due
to the charge carriers of the $\pi $ band, even the background
conductance quite often follows the ''semiconductive'' behavior,
namely, the slope of the $dV/dI$ curve at large biases is negative
(Fig. \ref{negative}). That means, that the carriers in the $\pi $
band are close to localization \cite{Kuz'menko}.

In the lower panel of Fig.\,\ref{MgB2hist} the theoretical
prediction of the energy gap distribution \cite{Choi} is shown.
One can see that the theoretical positions of the distribution
maxima coincide approximately with the experimental values. Only
the low-lying maximum is not seen in the experiment. It should be
noted that according to Mazin {\it et al.} \cite{Mazin2} variation
of the superconducting gaps inside the $\sigma$ and $\pi$ bands
can hardly be observed in real samples.

The distribution of the different gaps over the Fermi surface is
shown in Fig. \ref{Choi-D}. One can immediately see that for a
c-axis oriented film the main structure should have a smaller gap,
which is approximately isotropic. Only if the contact touches the
side face of a single crystallite (Fig.\,\ref{films}), is the
larger gap visible, since it corresponds to the cylindrical parts
of the Fermi surface with Fermi velocity parallel to the {\it ab}
plane.

\subsubsection{Single crystals}

\begin{figure}
\includegraphics[width=8cm,angle=0]{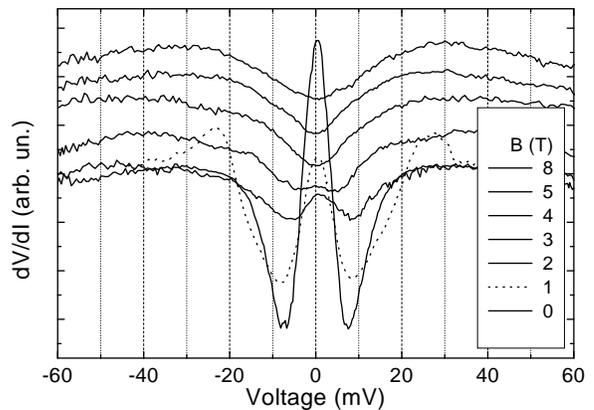}
\caption[]{Large gap structure evolution for single crystal
MgB$_2$-Au 87\,$\Omega$ junction in magnetic field at 4.2\,K. The
curves are shifted vertically for clarity.} \label{largegap}
\end{figure}

The same variety of energy gap structure is observed for single
crystals as well, but with some peculiarity due to preferential
orientation along the {\it ab} plane. The most amazing of them is
the observation of $dV/dI$-gap structure in Fig.\,\ref{largegap}
with visually only the larger gap present. This gap persist in a
magnetic field of a few tesla unlike the smaller gap, which
according to \cite{Szabo,Gonnelli} vanishes above 1\,T. Spectra of
that kind were not observed in thin films. This means that the
conductivity is governed only by the $\sigma $ band. This may be
caused by the circumstance that the $\pi $ band is blocked
completely by Mg disorder or by oxidation of Mg atoms on {\it
ab}-side surface of the crystal. At the same time, in a single
crystal there is much less scattering in the boron planes, due to
the robustness of the B-B bonds. We will see below that just this
case enables us to observe directly the most important $E_{2g}$
phonon mode in the electron-phonon interaction within the $\sigma$
band.

In single crystals the negative slope in $dV/dI$ curve at large
biases is observed quite often, which confirms that the disorder
in the $\pi $ band leads to quasi-localization of charge carriers.
An example of this is already shown in Fig. \ref{negative}.

\begin{figure}
\includegraphics[width=8.5cm,angle=0]{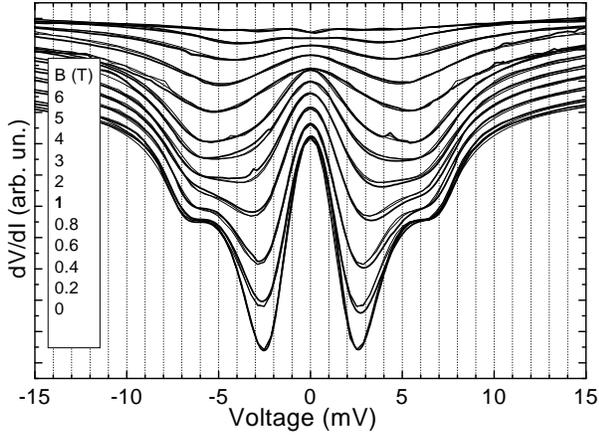}
\caption[]{Magnetic field dependences of $dV/dI$ curves (solid
lines) for a single crystal MgB$_2$-Cu 2.2$\Omega$ junction along
the $ab$ plane with their BTK fittings (thin lines). Two separate
sets of gap minima are clearly seen at low fields.}
\label{M163aa1}
\end{figure}

\begin{figure}
\includegraphics[width=8.5cm,angle=0]{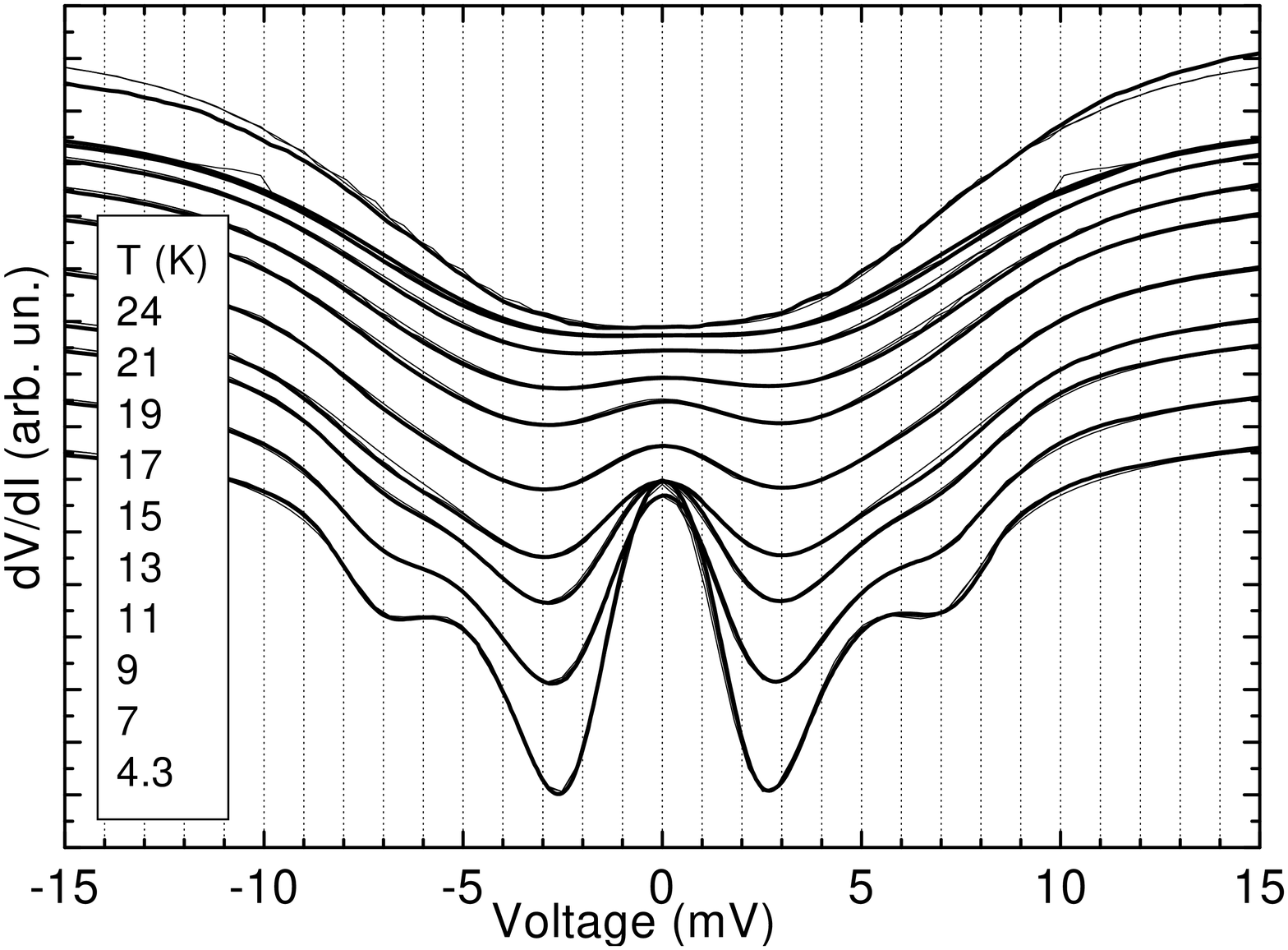}
\caption[]{Temperature dependences of $dV/dI$ curves (solid lines)
for the same junction as in Fig.\,\ref{M163aa1}  with their BTK
fittings (thin lines).} \label{M163aa2}
\end{figure}

\begin{figure}
\includegraphics[width=8.5cm,angle=0]{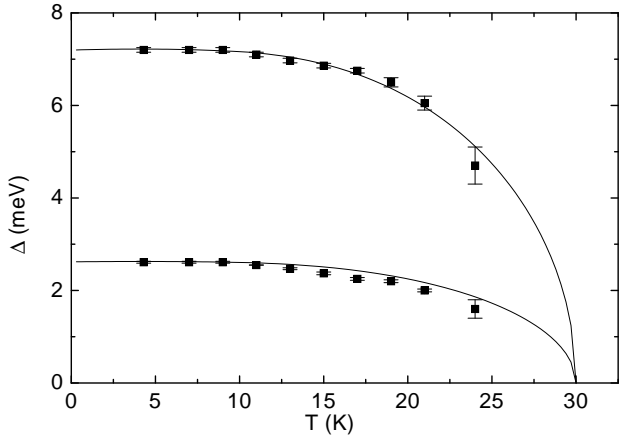}
\caption[]{Temperature dependences of large and small
superconducting energy gaps obtained by BTK fitting from
Fig.\,\ref{M163aa2}. The solid lines represent BCS-like behavior.}
\label{M163aa3}
\end{figure}

Figures \ref{M163aa1} and  \ref{M163aa2} display a series of
magnetic-field and temperature dependence of $dV/dI$ curves with
their BTK fit. Here the two gaps are clearly visible,
corresponding to the theoretical prediction in the {\it ab}
direction Eq.\,(\ref{brinkeq2}). The temperature dependence of
both gaps follows the BCS prediction (see Fig.\,\ref{M163aa3}).
For temperatures above 25\,K their behavior is unknown because
this particular contact did not survive the measurements likely
due to thermal expansion of the sample holder.

\begin{figure}
\includegraphics[width=8cm,angle=0]{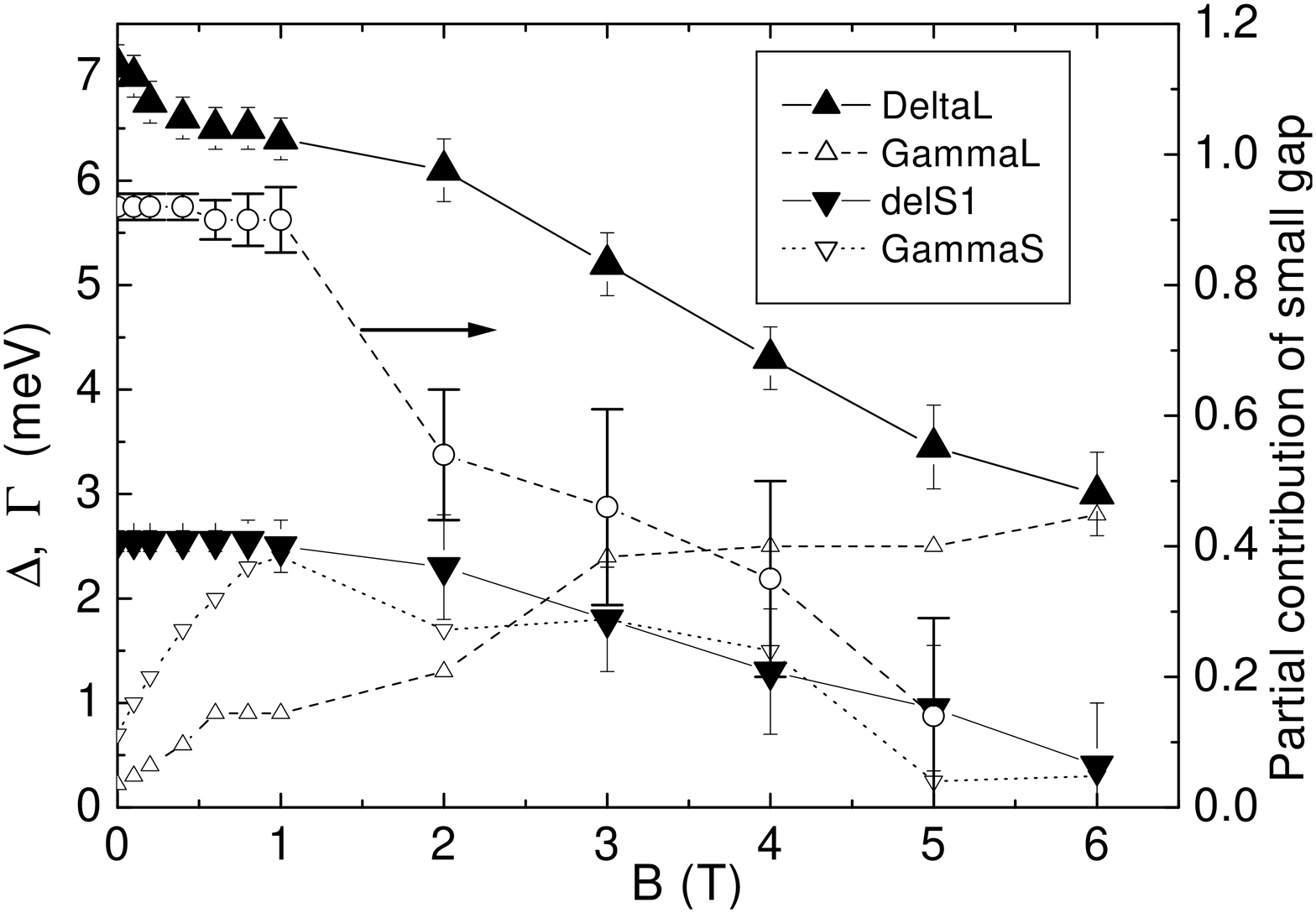}
\caption[]{Magnetic field dependences of the large and small
superconducting energy gaps (solid triangles) obtained by BTK
fitting from Fig.\,\ref{M163aa1}. Open triangles show the $\Gamma$
value for large and small gap, respectively. The circles
demonstrate the depression of the small-gap contribution to the
$dV/dI$ spectra by magnetic field. The lines connect the symbols
for clarity.} \label{deltH163}
\end{figure}

Figure \ref{deltH163} displays the magnetic field dependences of
large and small gaps. Surprisingly, the small gap value is not
depressed by a field of about 1\,T, and the estimated critical
field of about 6\,T is much higher as stated in
\cite{Gonnelli,Samuely}, although the intensity of the small-gap
minima is suppressed rapidly by a field of about 1\,T.
Correspondingly, the small-gap contribution $w$ \footnote{ $w$
inversely depends on the $\Gamma$ value, therefore the nearly
constant $w$ value between 0 and 1\,T is due to the fact that
$\Gamma$ rises by factor 4 at 1\,T} to the $dV/dI$ spectra is
decreased by magnetic field significantly, from 0.92 to 0.16 (see
Fig.\,\ref{deltH163}), while $w$ versus temperature even increases
slightly from 0.92 at 4.3\,K to 0.96 at 24\,K (not shown).

\begin{figure}
\includegraphics[width=8cm,angle=0]{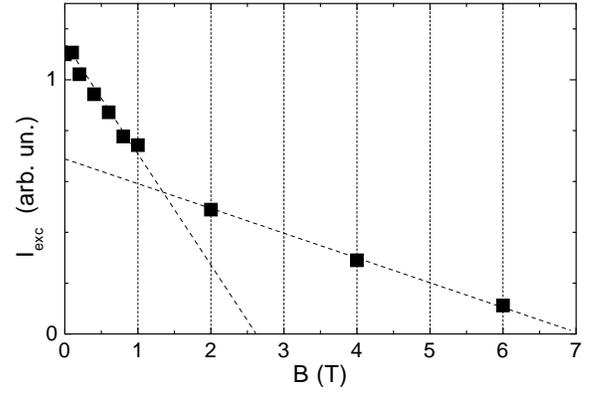}
\caption[]{$I_{exc}(B)$ (squares) for a MgB$_2$-Cu junction from
Fig.\,\ref{M163aa1}. The dashed lines show the different behavior
of $I_{exc}(B)$ at low and high fields. } \label{163-175b}
\end{figure}

The area under the energy-gap minima in $dV/dI(V)$ is
approximately proportional to the excess current $I_{exc}$ (see
Eq.(\ref{excess})) at $eV\gg\Delta$ (or roughly to the superfluid
density). The excess current depends on the magnetic field with a
positive overall curvature (Fig.\,\ref{163-175b}). $I_{exc}(B)$
decreases abruptly at first and then more slowly above 1\,T. This
corresponds to a drastic depressing of the $dV/dI(V)$
small-gap-minima intensity by a magnetic field of about 1\,T and
to robustness of the residual superconducting structure against
further increase of magnetic field. This is a quite different
dependence from what is expected for $I_{exc}$, which is in
general proportional to the gap value (\ref{excess}).

In contrast, $I_{exc}(T)$ has mostly negative curvature and shape
similar to the BCS dependence. Often a positive curvature appears
above 25\,K (see, e.g. Fig.\,\ref{onegap}).

\begin{figure}
\includegraphics[width=8cm,angle=0]{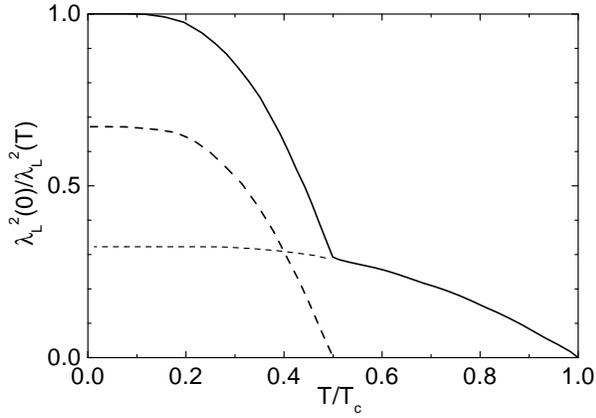}
\caption[]{Temperature dependence of the penetration depth in the
model of two independent BCS superconducting bands (dashed and
dotted line) with different superconducting gaps. The resulting
penetration depth (solid line) clearly shows a non-BCS temperature
behavior. The low-temperature behavior will be dominated by the
band with the smaller superconducting gap. After Golubov \etal
\cite{Golubov2}.} \label{pen_dept}
\end{figure}

\begin{figure}
\includegraphics[width=8.5cm,angle=0]{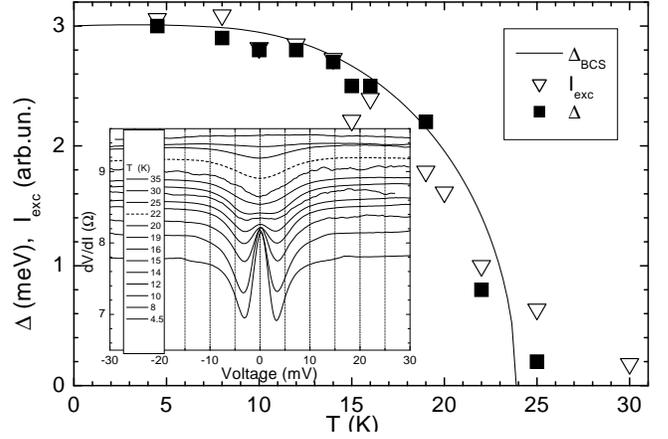}
\caption[]{Temperature dependence of a single superconducting
energy gap (squares) obtained by BTK fitting of $dV/dI$ curves
from inset. The solid lines represent BCS-like behavior. The
triangles show the dependence of the excess current. Inset:
$dV/dI$ curves for a MgB$_2$-Cu 8\,$\Omega$ contact at different
temperatures.} \label{onegap}
\end{figure}

This kind of anomaly can be due to the two-band nature of
superconductivity in MgB$_2 $, since the magnetic field
(temperature) suppresses the superconductivity more quickly in the
$\pi $ band and then, at higher field (temperature), in the
$\sigma $ band. The same consideration is valid for $1/\lambda_L$,
which is roughly proportional to the "charge density of superfluid
condensate". In the case of zero interband scattering, the simple
model \cite{Golubov2} predicts the temperature dependence shown in
Fig. \ref{pen_dept} for $\sigma$ and $\pi$ parallel channels,
which will yield a smooth curve with general positive curvature,
taking into account the small interband scattering occurring in
reality.

If the $\pi $-band conductivity is blocked by a short mean free
path, then the curvature of $I_{exc}(T)$, being proportional to
$\Delta _\sigma (T)$, should be negative, which supplies us with
additional confirmation of single band conductivity along the
$\sigma$ band. Thus, measuring the magnetic field and temperature
dependences of $I_{exc}$ can elucidate the contact structure.

Figure \ref{onegap} displays the temperature dependence of the gap
for the $dV/dI$ curves with a single gap structure, which vanishes
around 25\,K. A magnetic field of 1\,T suppresses the gap minima
intensity by factor of two, but the minima are clearly seen even
at 4\,T (not shown), the maximal field in this experimental trial.
This excludes an origin of these gap minima due to small gap.
 According to the calculation in \cite{Brinkman} a large
amount of impurity scattering will cause the gaps to converge to
$\Delta \simeq$4.1\,meV and $T_c$ to 25.4\,K. Therefore these
single-gap spectra reflect a strong interband scattering due to
impurities, which likely causes a "semiconducting-like" behavior
of $dV/dI$ above T$_c$ (see Fig.\,\ref{onegap}, inset).
$I_{exc}(T)$ behaves nearly as $\Delta(T)$ except in the region
T$>$25\,K, where $I_{exc}$ is still nonzero because of a residual
shallow zero-bias minimum in $dV/dI$ above 25\,K.

\subsection{Phonon structure in the $I-V$ characteristics}

\subsubsection{PC EPI spectra of nonsuperconducting diborides}

\begin{figure}
\includegraphics[width=8.5cm,angle=0]{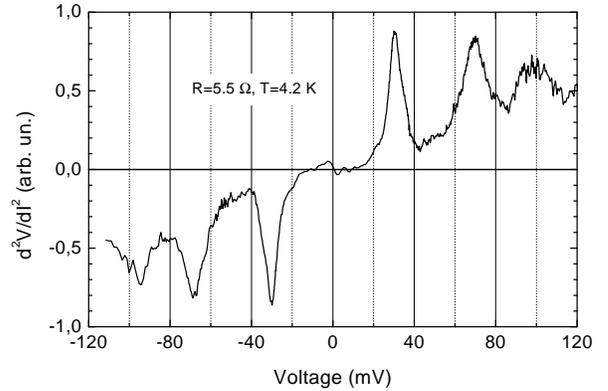}
\caption[]{Raw PC EPI spectrum for a ZrB$_2$ 5.5\,$\Omega$ point
contact at 4.2\,K. After Naidyuk \etal \cite{Naidyuk2}.}
\label{MeBf1}
\end{figure}

We have studied the PC EPI spectra  $ d^2V/dI^2 \propto
-d^2I/dV^2$ (see also Eq.\,(\ref{pcs})) of non-superconducting
diborides MeB$_2$ (Me=Zr, Nb, Ta) \cite{Naidyuk2}. The cleanest
sample we have is a ZrB$_2$ single crystal, and its PC EPI
spectrum is shown in Fig.\,\ref{MeBf1}. One recognizes a classical
PC EPI spectrum from which one can estimate the position of 3 main
phonon peaks and obtain the lower limit of EPI parameter $\lambda
_{PC}$ \cite{Naidyuk2}.

\begin{figure}
\includegraphics[width=8.5cm,angle=0]{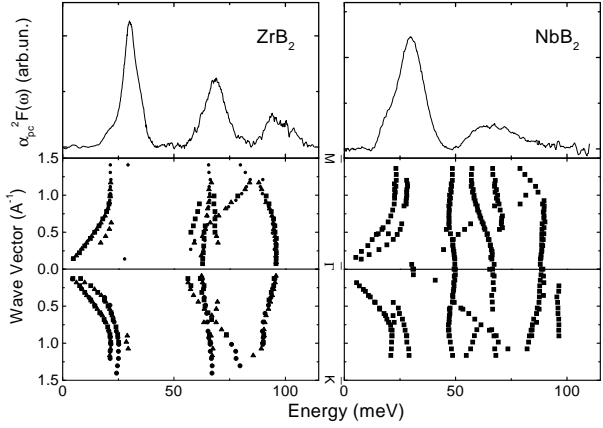}
\caption[]{Comparison of high-resolution electron-energy-loss
spectroscopy measurements of surface phonon dispersion (bottom
panels, symbols) \cite{Aizawa} with the PC spectra for ZrB$_2$ and
NbB$_2$ after subtraction of the rising background (upper
panels).} \label{Aizawa2}
\end{figure}

\begin{figure}
\includegraphics[width=7cm,angle=0]{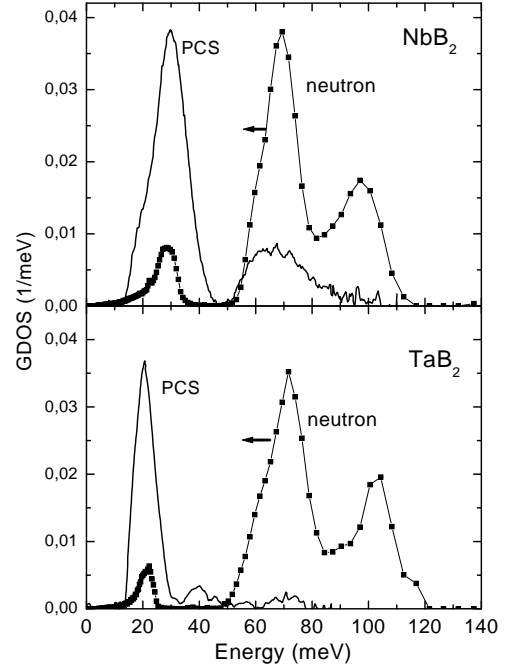}
\caption[]{Comparison of phonon DOS neutron measurements after
Heid \etal \cite{Heid} (symbols) with PC spectra for NbB$_2$ and
TaB$_2$ after subtraction of the rising background (solid
curves).} \label{Heid}
\end{figure}

Essentially similar spectra were observed for other diborides,
taking into account their purity and increased EPI, which leads to
a transition from the spectroscopic to a non-spectroscopic
(thermal) regime of current flow \cite{Naidyuk2}. The positions of
the low-energy peaks are proportional to the inverse square root
of the masses of the $d$ metals \cite{Naidyuk2}, as expected. For
these compounds the phonon density of states is measured by means
of neutron scattering \cite{Heid} and the surface phonon
dispersion is derived by high-resolution electron-energy-loss
spectroscopy \cite{Aizawa}. The positions of the phonon peaks or
$d\omega/dq$=0 for the dispersion curves correspond to maxima of
the PC spectra (Fig.\,\ref{Aizawa2}, Fig.\,\ref{Heid}).

\subsubsection{PC EPI spectra of MgB$_2$ in {\it c}-axis oriented films}

\begin{figure}
\includegraphics[width=8cm,angle=0]{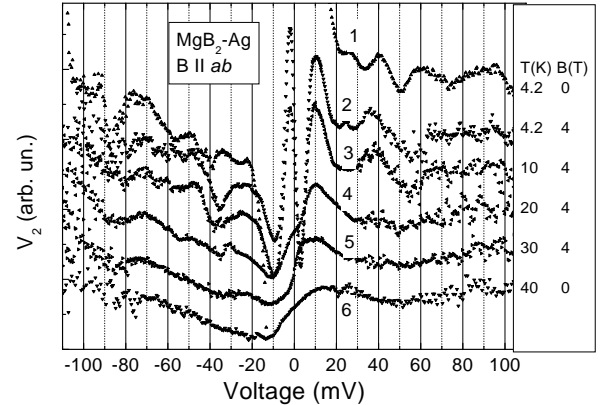}
\caption[]{Phonon singularities in the PC spectra of a MgB$_2$
thin-film -- Ag junction as a function of magnetic field and
temperature. $T$ and $ B$ are shown beside each curve. After
Yanson \etal \cite{YansonPRB}. } \label{Fig4a}
\end{figure}

From the above consideration we had anticipated that one could
easily measure the EPI spectral function of MgB$_2$ in the normal
state, provided that the superconductivity is destroyed by
magnetic field. Unfortunately, that was not the case. The stronger
we suppress the superconductivity in MgB$_2$, the less traces of
phonon structure remain in the $I-V$ characteristic and its
derivatives (Fig.\,\ref{Fig4a}) \cite{YansonPRB}. This is in odd
in relation to the classical PCS, since the {\it inelastic} phonon
spectrum should not depend on the state of electrodes in the first
approximation (see section {\bf Theoretical background}).

\begin{figure}
\includegraphics[width=8cm,angle=0]{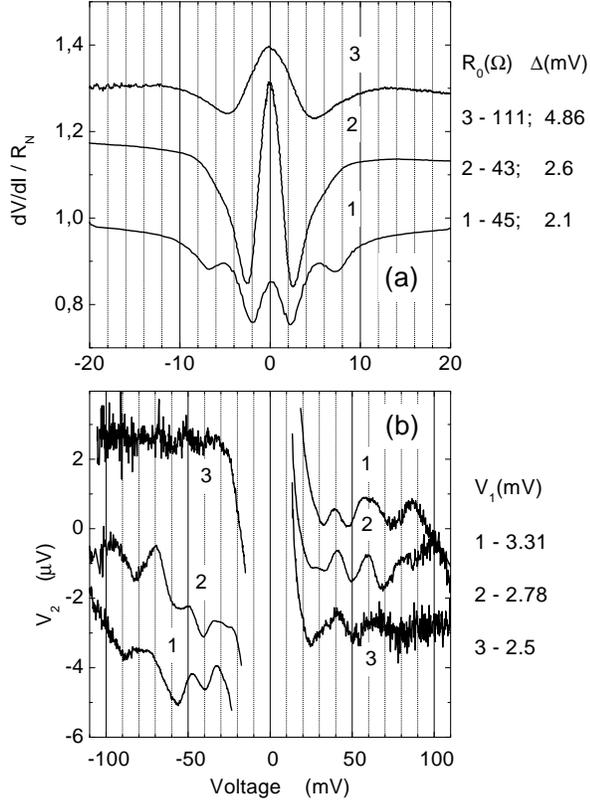}
\caption[]{Superconducting gap features (upper panel) and phonon
structure (bottom panel) in the spectra of thin-film MgB$_2$-Ag
junctions with different resistances at $T=4.2 K$, $B=0$. After
Yanson \etal \cite{YansonPRB}.} \label{Fig2phon}
\end{figure}

Instead, most of the MgB$_2$ spectra in the superconducting state
show reproducible structure in the phonon energy range
(Fig.\,\ref{Fig2phon}) which was not similar to the expected
phonon maxima superimposed on the rising background. This
structure disappears upon transition to the normal state. Quite
interestingly is that the intensity of this structure increases
with increase of the value of the small gap, which means that the
gap in the $\pi $ band and observed phonon structure is connected
\cite{YansonPRB}. Based on the theoretical consideration mentioned
above, we conclude that the disorder in the $\pi $ band is so
strong that it precludes observation of the {\it inelastic
current}, and the phonon nonlinearities of the excess current (see
Eq.(\ref{Om-Kul-Bel})) play the main role, which does not depend
on the scattering.

\begin{figure}
\includegraphics[width=7.5cm,angle=0]{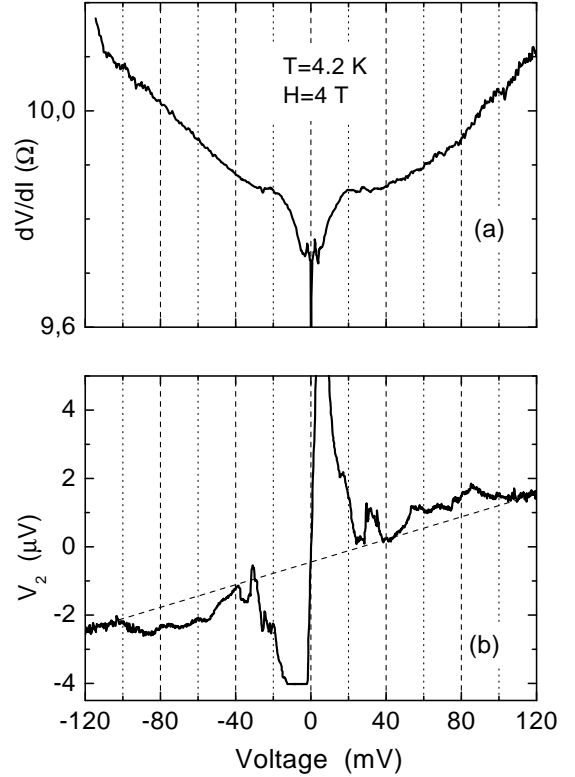}
\caption[]{$dV/dI$ and $V_2\propto d^2V/dI^2$ curves for a thin
film MgB$_2$-Ag junction revealing the inelastic PC spectrum for
the $\pi$ band. After Bobrov \etal \cite{Bobrov}. } \label{Bobrov}
\end{figure}

\begin{figure}
\includegraphics[width=7.5cm,angle=0]{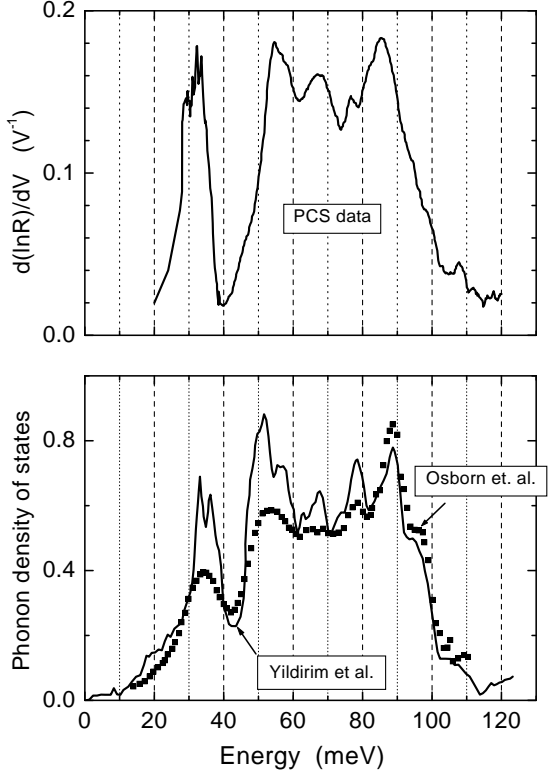}
\caption[]{Comparison of the PC EPI spectrum (upper panel) from
Fig.\,\ref{Bobrov} (after subtraction of the linear background and
zero-bias maxima below 25 meV) with the phonon DOS measured by
neutron scattering \cite{Yildirim,Osborn} (bottom panel). }
\label{Bobrov1}
\end{figure}

\begin{figure}
\includegraphics[width=8.5cm,angle=0]{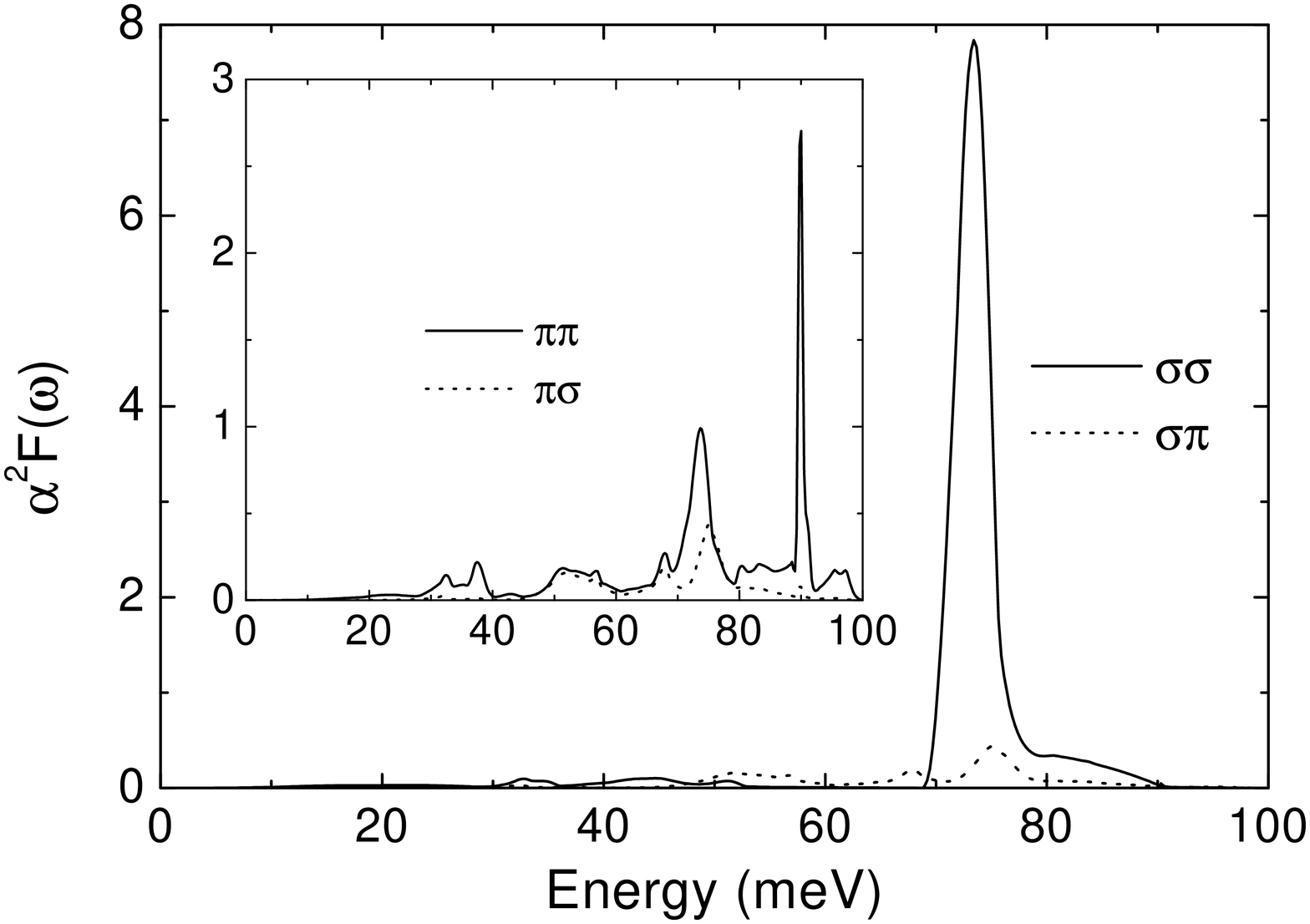}
\caption[]{Calculated Eliashberg functions for the $\sigma$ and
$\pi$ bands (inset). After Golubov \etal \cite{Golubov}. }
\label{Golubov}
\end{figure}

Very rarely we observed signs that the measured characteristics
indeed satisfy the conditions imposed on the inelastic PC spectra.
One such example is shown in Figs.,\ref{Bobrov}, \ref{Bobrov1}.
For this particular junction the superconducting peculiarities are
almost completely suppressed above 20\,mV by moderate field
(4\,T). What remains is a weak zero-bias minimum ($\sim1$\%) from
the rather high value of the gap (Fig.\,\ref{Bobrov}). The
background in $dV/dI(V)$ rises nearly quadratically up to a few \%
of $R_0$ at large biases ($\sim $100 mV). This leads to a linear
background in $V_2\propto d^2V/dI^2(V)$ with phonon peaks
superposed above the background both in negative and positive bias
polarity (compare with Fig.\,\ref{MeBf1}). The structure observed
above 30\,meV corresponds reasonably in shape to the phonon
density of states (Fig.\,\ref{Bobrov1}). At the low voltages
(below 30 meV), most probable, the gap peculiarities still prevail
over the $d^2V/dI^2(V)$ structure. Thus, for this contact we
assume to observe the {\it inelastic} PC spectrum for the $\pi$
band, which should be compared to the Eliashberg EPI function for
the same band calculated in Ref. \cite{Golubov}
(Fig.\ref{Golubov}). Both the experimental spectrum and the
$\pi$-band Eliashberg function do not show anomalously high
intensity of $E_{2g}$ phonon mode, since only the Eliashberg
function for $\sigma$ band is the principal driving force for high
$T_c$ in MgB$_2$. The same conclusion should be ascribed to the
excess-current phonon structure, since it also corresponds to the
$\pi $ band. This band has much larger Fermi velocity and plasma
frequency along the {\it c}-axis compared to the $\sigma $ band
\cite{Brinkman}.

Thus, in order to register the principal EPI with the $E_{2g}$
phonon mode, we are faced with the necessity of measuring the PC
spectra for only the $\sigma$  band. This can be done in a single
crystal along the {\it ab} plane with blocked $\pi-$ band
conductivity.

\subsubsection{PC EPI spectra of MgB$_2$ in the {\it ab} direction}

\begin{figure}
\includegraphics[width=7.5cm,angle=0]{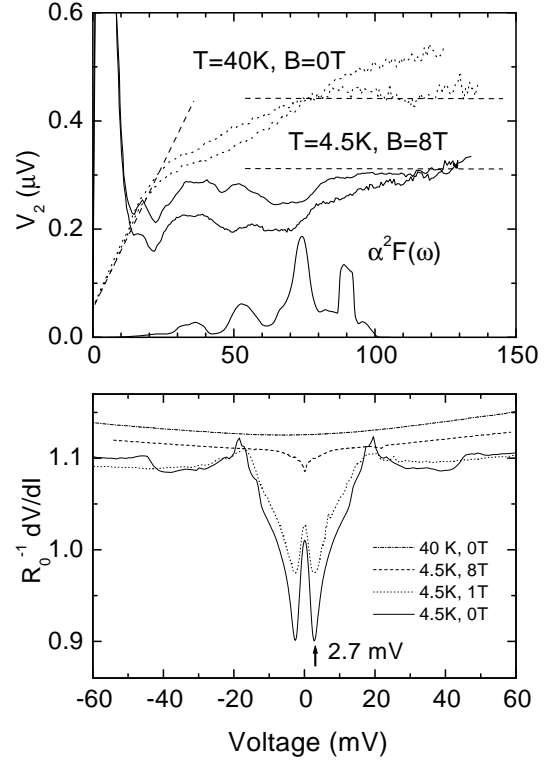}
\caption[]{ $V_2\propto d^2V/dI^2$ (for two bias voltage
polarities) and $dV/dI$ curves for a single crystal MgB$_2$-Cu
junction ($R_0=1.5\,\Omega$) along the $ab$ plane. Here the
conductivity along the $\pi$ band prevails, as is shown by the
pronounced small-gap structure for the zero-field $dV/dI$ curve at
4.5\,K. The $\alpha^2 F(\omega)$ curve is the theoretical
prediction for the $\pi$-band Eliashberg function from
Fig.\,\ref{Golubov} (inset) smeared similarly to the experimental
data. After Naidyuk \etal \cite{Naidyuk3}.} \label{Fig2Search}
\end{figure}

The desired situation was described in Ref.\,\cite{Naidyuk3} for
single crystal oriented in the {\it ab} plane. As was mentioned
above, the nominal orientation of the contact axis to be parallel
to {\it ab} plane is not enough to be sure that this situation
occurs in reality. Moreover, even if one establishes the necessary
orientation (i.\,e., contact axis parallel to {\it ab} plane) the
spectra should reflect both bands with a prevalence of the
undesired $\pi $ band, because due to spherical spreading of the
current the orientational selectivity of metallic point contact is
much worse than that for the plane tunnel junction, where it goes
exponentially. The large mixture of $\pi $-band contribution is
clearly seen from the gap structure in Fig.\,\ref{Fig2Search}.
Inside the wings at the biases corresponding to the large gap
(supposed to belong to the $\sigma $-band gap) the deep minima
located at the smaller gap (correspondingly to the $\pi $-band
gap) are clearly seen (see bottom panel of
Fig.\,\ref{Fig2Search}). The EPI spectrum of the same junction is
shown in the upper panel. One can see that the nonlinearities of
the $I-V$ characteristic at phonon biases are very small, and a
reproducible structure roughly corresponding to the Eliashberg EPI
function of the $\pi $ band \cite{Golubov,Dolgov} appears in the
bias range 20 $\div$ 60\,mV. Above 60 mV the PC spectrum broadens
sufficiently hidden higher-lying phonon maxima. Even in the normal
state ($T\geq T_c$), where the excess current disappears, one can
see the kink at $\approx 20\div 30$ meV, where the first peak of
the phonon DOS and the first maximum of the Eliashberg EPI
function of the $\pi $ band occurs.
At eV$\approx 90\div 100$ meV the PC EPI spectrum of Fig.
\ref{Fig2Search} saturates just where the phonon DOS ends. At
$T\geq T_c$ intermediate phonon peaks are hardly seen, since the
thermal resolution, which equals 5.44\,$k_BT$, amounts about
20\,meV, and the regime of current flow is far from ballistic, due
to the high background observed. No prevalence of the $E_{2g}$
phonon mode is observed, like a big maximum of EPI at $\approx
60\div 70$\,meV or a kink at $T\geq T_c$ for these biases.

\begin{figure}
\includegraphics[width=7.5cm,angle=0]{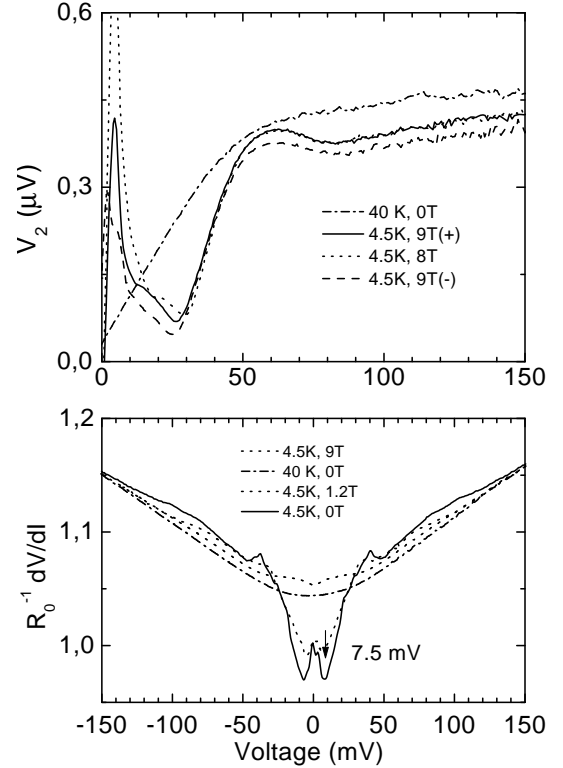}
\caption[]{$V_2\propto d^2V/dI^2$ (for two bias voltage polarities
at 9\,T) and $dV/dI$ curves for a single-crystal MgB$_2$-Cu
junction ($R_0=7.2\,\Omega$) along the $ab$ plane. Here the
conductivity along the $\sigma$ band prevails, as is shown by
pronounced large-gap structure for the zero field $dV/dI$ curve at
4.5\,K. After Naidyuk  \etal \cite{Naidyuk3}. } \label{Fig1Search}
\end{figure}

A quite different spectrum is shown in Fig.\,\ref{Fig1Search},
which is our key result. Consider first the $dV/dI(V)$
characteristics (see bottom panel). The energy gap structure shows
the gap minima corresponding to the large gap ($ \sigma $-band
gap). The increase of $dV/dI(V)$ at larger biases is noticeably
larger than in the previous case (Fig.\,\ref{Fig2Search}). One can
notice that the relatively small magnetic field ($\sim $1 T) does
not decrease the intensity of gap structure substantially, unlike
those for Fig. \ref{M163aa1}, and even less than for
Fig.\,\ref{Fig2Search}. According to \cite{Szabo,Gonnelli} a field
of about 1\,T should depress the small gap intensity completely.
All these facts evidence that we obtain a contact, in which only
the $\sigma $-band channel in conductivity is operated.

Let us turn to the PC\ EPI spectra $d^2V/dI^2(V)$, which are
connected via the following expression to the second harmonic
signal $V_2$ recorded in experiment:
\[
\frac 1{R^2}\frac{d^2V}{dI^2}=2\sqrt{2}\frac{V_2}{V_1^2}
\]
Here $R=dV/dI$ and $V_1$ is the rms value of the modulation
voltage for the standard technique in the tunneling and
point-contact spectroscopy.

The PC EPI spectra for this contact are shown in
Fig.\,\ref{Fig1Search} (upper panel) for the highest field
attainable in our experiments \cite{Naidyuk3}. One can see that
8$\div 9$ T is still not enough to destroy completely the
superconductivity in the energy-gap low bias range (0$\div 30$
meV), which can be taken as characteristic for a strongly
superconducting $\sigma $ band. On the other hand, at larger
biases no influence of field is noted, which evidences that this
part of $I-V$ characteristic does not contains superconducting
peculiarities, likely due to the high current density in the
contact. Except for a small asymmetry, the spectrum is reproduced
for both polarities. Before saturation at biases $\geq 100$ meV,
where the phonon DOS ends, a well-resolved wide bump occurs, which
is located at about 60\,meV. Further on, we will concentrate on
this.

\begin{figure}
\includegraphics[width=8.5cm,angle=0]{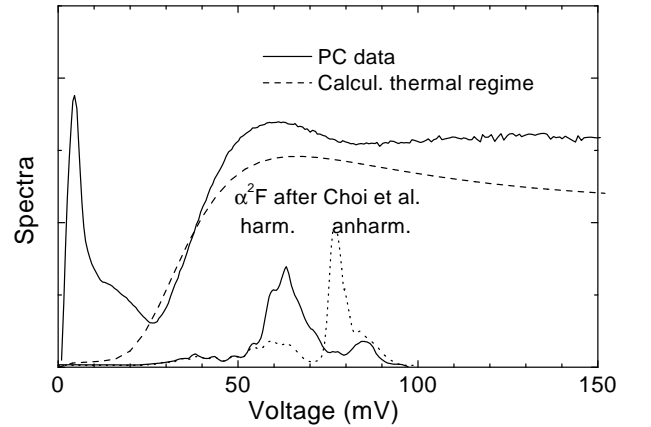}
\caption[]{Comparison of the experimental spectrum of
Fig.\,\ref{Fig1Search} with the thermal spectrum for a model
spectral function in the form of a Lorentzian at 60\,meV with a
width of 2\,meV (dashed line) and with the theoretical EPI spectra
(bottom curves). After Naidyuk  \etal \cite{Naidyuk3}. }
\label{Fig4Search}
\end{figure}

\begin{figure}
\includegraphics[width=7.5cm,angle=0]{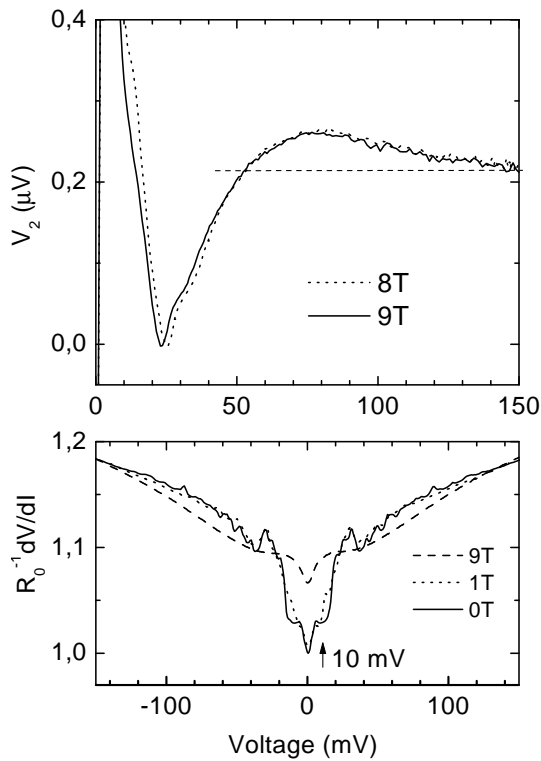}
\caption[]{Thermal limit for $\sigma$ band (as shows expressed
large gap structure for zero field $dV/dI$ curve at 4.5\,K) in PC
spectrum of MgB$_2$ single crystal along $ab$ plane. After Naidyuk
\etal \cite{Naidyuk3}.} \label{Fig3Search}
\end{figure}

First, we rescale it to the spectrum in $R_0^{-1}dR/dV$ units, in
order to compare with the theoretical estimation. We will show
that the bump is of spectroscopic origin, i.\,e. is the regime of
current flow through the contact is not thermal, although the
background at large biases ($V\geq 100$ meV) is high. To do so, we
compare this bump with a PC spectrum in the thermal regime for a
model EPI function, which consists of a Lorentzian at 60\,meV with
small (2 meV) width. Calculated according to Kulik
\cite{Kulik_therm}, the thermal PC\ EPI spectrum is much broader,
shown in Fig.\,\ref{Fig4Search} as a dashed line.
Any further increase of the width of the model spectra will
broaden the curve obtained. Comparing the experimental and model
spectra enable us to conclude that in spite of the large width,
the maximum of the experimental spectra still corresponds to the
spectroscopic regime. The high-temperature ($T\geq T_c$) spectrum
in Fig.\,\ref{Fig1Search} shows the smeared kink at about 60\,meV,
unlike that of Fig.\,\ref{Fig2Search}. Introducing greater
disorder in the boron plane by a fabrication procedure or by
trying other spots on the side-face surface, the smeared thermal
spectra were observed, coinciding in shape with the dashed curve
in Fig.\,\ref{Fig3Search}. In this figure another junction is
shown, where the energy gap structure also points to the $\sigma
$-band channel. Other junctions display the kink at about 30$\div
40$ meV, like the high temperature spectrum in
Fig.\,\ref{Fig2Search}, which together with their energy-gap
structure can be ascribed to the thermal limit mainly in the $\pi$
band, despite the rather low bath temperature.

A PC spectrum with broad maxima including also one at about 60 mV
were observed in \cite{Samuely} on polycrystalline MgB$_2$ samples
driven to the normal state by applying moderate magnetic field and
increasing of the temperature.

The large width of the EPI peak connected with the $E_{2g}$ phonon
mode (Fig.\,\ref{Fig1Search}) is not surprising. Shukla {\it et
al.} \cite{Shukla} measured the phonon dispersion curves along the
$\Gamma $A and $ \Gamma $M directions by means of inelastic X-ray
scattering (see Fig.\,\ref{Shukla1}). The full width at half
maximum for the $E_{2g}$ mode along the $\Gamma $A direction
amounts about 20-28 meV, which corresponds well to what we observe
in the point-contact spectrum. If the phonon lifetime corresponds
to this (inverse) energy, then the phonon mean free path is about
equal to the lattice constant \cite{Naidyuk3}, and due to phonon
reabsorption by accelerating electrons, we should anticipate a
large background in the PC spectra as observed. If we compare the
position of the bump ($\approx 60$ meV) with what is predicted for
isotropic Eliashberg EPI function \cite{ChoiPRB} (see
Fig.\,\ref{Fig4Search}), then we, together with Shukla {\it et
al.,} should admit that the phonon-phonon anharmonicity is
inessential for this mode, and its high width is due completely to
the EPI.

Now turn to the nonlinearity of the $I-V$ curves due to
electron-phonon interaction, which can be estimated from the
$dV/dI$ curves as about 10\% for contact with the E$_{2g}$ phonon
modes in Fig.\,\ref{Fig1Search}. This is comparable with the
nonlinearity observed for nonsuperconducting diborides
\cite{Naidyuk2} with a small electron-phonon coupling constant.
The reason for the relatively low nonlinearity of the $I-V$ curves
and low intensity of the principal E$_{2g}$ phonon modes in the
spectra for the MgB$_2$ contacts can be the fact that anomalous
strong interaction is characteristic for restricted group of
phonons with sufficiently small wave vector \cite{Mazin1}, whereas
in point-contact spectroscopy the large angle scattering is
underlined.

\section{Conclusions}

We made an overview of the PCS investigations of $c$-axis oriented
thin films and single crystals of MgB$_2$. Our conclusions are as
follows:

1. There are two different superconducting gaps in MgB$_2$, which
are grouped at 2.4 and 7.0 meV. Roughly, in half of all point
contacts studied for $c$-axis oriented films the two gap structure
merges together due to strong elastic scattering remaining a
single gap at about 3.5 meV.

2. Anomalous temperature and especially magnetic field
dependencies of excess current in point-contact junctions reflect
the two-band structure of the superconducting order parameter in
MgB$_2$.

3. There are two mechanisms of revealing phonon structure in the
PC spectra of MgB$_2$: i) through the inelastic backscattering
current, like for ordinary point-contact spectroscopy, and ii)
through the energy dependence of the excess current, like in the
similar tunneling spectroscopy of the electron-phonon interaction.
They can be discriminated by destroying the superconductivity with
a magnetic filed and/or temperature, and by varying the electron
mean free path.

4. The prevailing appearance of the $E_{2g}$ boron mode, which
mediates the creation of Cooper pairs, is seen in the PC spectra
only along the $a-b$ direction in accordance with the theory. The
relatively small intensity of this mode in the PC spectra is
likely due to their small wave vector and restricted phase volume.

5. Related diborides (ZrB$_2$, NbB$_2$, and TaB$_2$) have
$d^2V/dI^2$  spectra proportional to the electron-phonon
interaction spectral function like that in common metals and a
small EPI constant corresponding to their nonsuperconducting
state.

\section*{Acknowledgements}
The authors are grateful to N. L. Bobrov, P. N. Chubov, V. V.
Fisun, O. E. Kvitnitskaya, and L. V. Tyutrina for collaboration
during the MgB$_2$ investigation. IKY thanks Institute of Solid
State Physics in Forschungzentrum Karlsruhe for hospitality, and
Prof. H. von L\"{o}hneysen for constant support. The work in
Ukraine was supported by the State Foundation of Fundamental
Research under Grant $\Phi$7/528-2001.


\section*{Note added in proof}

After the paper was completed we have learned of the paper by
Koshelev and Golubov \cite{Koshelev}, where the magnetic field
dependence of $\Delta _\sigma $ and $\Delta _\pi$ was presented.
It turned out that the $\Delta _\sigma (B)$ and $\Delta _\pi (B)$
behavior is different and is governed by diffusion constants
depending on the coherence length. However, the critical field is
the same both for $\Delta _\sigma $ and $\Delta _\pi $. This is in
line with our observation given in Fig.\,\ref{deltH163}.
Additionally, two experimental reports on the effect of magnetic
field on both gaps in MgB$_2$ by Gonnelli \etal
~(cond-mat/0308152) and Bugoslavsky \etal ~(cond-mat/0307540)
appeared in the E-print archive. Bugoslavsky \etal ~reported that
both order parameters persist to a common magnetic field. Gonnelli
\etal ~corrected their previous claims and mentioned that
identification of the magnetic field at which the $\pi$-band
features in $dV/dI$ visually disappear with the critical field for
the $\pi$  band might not be correct.



\begin{thebibliography}{}
\bibitem{Nagamatsu}  Jun Nagamatsu, Norimasa Nakagawa, Takahiro Muranaka,
Yuji Zenitani, Jun Akimitsu, Nature {\bf 410}, 63 (2001).

\bibitem{Mazin}  I.I. Mazin, O.K. Andersen, O. Jepsen, O.V. Dolgov, J.
Kortus, A.A. Golubov, A.B. Kuz'menko, and D. van der Marel, Phys.
Rev. Lett. {\bf 89}, 107002 (2002).

\bibitem{Gurevich}  A. Gurevich, Phys. Rev. B {\bf 67}, 184514 (2003).

\bibitem{An}  J.M. An and W.E. Pickett, Phys. Rev. Lett. {\bf 86}, 4366
(2001).

\bibitem{Kong}  Y. Kong, O.V. Dolgov, O. Jepsen, and O.K. Andersen, Phys.
Rev. B {\bf 64}, 020501(R) (2001).

\bibitem{Kortus}  J. Kortus, I.I. Mazin, K.D. Belashchenko, V.P. Antropov,
and L.L. Boyer, Phys. Rev. Lett {\bf 86}, 4656 (2001).

\bibitem{Liu}  Amy Y. Liu, I.I. Mazin, and Jens Kortus, Phys. Rev. Lett.
{\bf 87}, 087005 (2001).

\bibitem{Yildirim}  T. Yildirim, O. G\"{u}lseren, J.W. Lynn, C.M. Brown,
T.J. Udovic, Q. Huang, N. Rogado, K.A. Regan, M.A. Hayward, J.S.
Slusky, T. He, M. K. Haas, P. Khalifah, K. Inumaru, and R.J. Cava,
Phys. Rev. Lett. {\bf 87}, 037001 (2001).

\bibitem{Mazin1}  I.I. Mazin, V.P. Antropov, Physica C {\bf 385}, 49
(2003).

\bibitem{Eltsev}Yu. Eltsev, K. Nakao, S. Lee, T. Masui, N. Chikumoto,
S. Tajima, N. Koshizuka, M. Murakami, Phys. Rev. B {\bf 66},
180504(R) (2002).

\bibitem{Brinkman}  A. Brinkman, A.A. Golubov, H. Rogalla, O.V.
Dolgov, J. Kortus, Y. Kong, O. Jepsen, and O.K. Andersen, Phys.
Rev. B {\bf 65}, 180517 (2002).

\bibitem{Kogan} V.G. Kogan, S.L. Bud'ko, Physica C {\bf 385}, 131
(2003).

\bibitem{Osborn}  R. Osborn, E.A. Goremychkin, A.I. Kolesnikov, and D.G.
Hinks, Phys. Rev. Lett. {\bf 87}, 017005 (2001).

\bibitem{Shukla}  Abhay Shukla, Matteo Calandra, Matteo d'Astuto, Michele
Lazzeri, Francesco Mauri, Christophe Bellin, Michael Krisch, J.
Karpinski, S.M. Kazakov, J. Jun, D. Daghero, and K. Parlinski,
Phys. Rev. Lett. {\bf 90}, 095506 (2003).

\bibitem{Bohnen}  K.-P. Bohnen, R. Heid, and B. Renker, Phys. Rev. Lett.
{\bf 86}, 5771 (2001).

\bibitem{Quilty}  James William Quilty, Physica C {\bf 385}, 264 (2003).

\bibitem{Goncharov}  A.F. Goncharov, V.V. Struzhkin, Physica C {\bf 385},
117 (2003).

\bibitem{Choi}  Hyoung Joon Choi, David Roundy, Hong Sun, Marvin L. Cohen,
Steven G. Louie, Nature {\bf 418}, 758 (2002).

\bibitem{ChoiPRB}  Hyoung Joon Choi, David Roundy, Hong Sun, Marvin L.
Cohen, and Steven G. Louie, Phys. Rev. B {\bf 66}, 020513(R)
(2002).

\bibitem{An1}  J.M. An, S.Y. Savrasov, H. Rosner, and W.E. Pickett, Phys.
Rev. B {\bf 66}, 220502(R) (2002).

\bibitem{Cappelluti} E. Cappelluti, S. Ciuchi, C. Grimaldi, L. Pietronero,
and S. Str\"{a}ssler, Phys. Rev. Lett. {\bf 88}, 117003 (2002).

\bibitem{Bouquet}  F. Bouquet ,Y. Wang, I. Sheikin, P. Toulemonde, M.
Eisterer, H.W. Weber, S. Lee, S. Tajima, A. Junod, Physica C {\bf
385}, 192 (2003).

\bibitem{YansonPRB}  I.K. Yanson, V.V. Fisun, N.L. Bobrov, Yu.G. Naidyuk,
W.N. Kang, Eun-Mi Choi, Hyun-Jung Kim, and Sung-Ik Lee, Phys. Rev.
B {\bf 67}, 024517 (2003).

\bibitem{Kang}  W. N. Kang, Hyeong-Jin Kim, Eun-Mi Choi, C.U. Jung, and
Sung-Ik Lee, Science {\bf 292}, 1521 (2001).

\bibitem{Sung-Ik}  W.N.Kang, Eun-Mi Choi, Hyeong-Jin Kim, Hyun-Jung Kim,
Sung-Ik Lee, Physica C {\bf 385}, 24  (2003).

\bibitem{Lee}  Sergey Lee, Physica C {\bf 385}, 31  (2003);
S. Lee, H. Mori, T. Masui, Yu. Eltsev, A. Yamanoto and S. Tajima,
J. Phys. Soc. of Japan, {\bf 70}, 2255 (2001).

\bibitem{Quilty1}  J.W. Quilty, S. Lee, A. Yamamoto, S. Tajima, Phys. Rev.
Lett. {\bf 88}, 087001(2001).

\bibitem{Rowell} J.M. Rowell, Supercond. Sci. Technol. {\bf 16},
R17-27 (2003).

\bibitem{Omel}  V.A. Khlus and A.N. Omel'yanchuk, Sov. J. Low Temp. Phys.
{\bf 9}, 189 (1983).

\bibitem{Khlus}  V.A. Khlus, Sov. J. Low Temp. Phys. {\bf 9}, 510 (1983).

\bibitem{KOS}  I.O. Kulik, A.N. Omelyanchouk, and R.I. Shekhter, Sov. J.
Low Temp. Phys. {\bf 3, }840 (1977).

\bibitem{YansonSC}  I.K. Yanson, in book I.O. Kulik and
R.Ellialtioglu (eds.){\it , Quantum Mesoscopic Phenomena and
Mesoscopic Devices in Microelectronic,} p. 61-77, (2000), Kluwer
Academic Publishers (see also: cond-mat/0008116).

\bibitem{Om-Kul-Bel}  A.N. Omel'yanchuk, S.I. Beloborod'ko, and I.O.
Kulik, Sov. J. Low Temp. Phys. {\bf 14}, 630 (1988).

\bibitem{Bel-Om}  S.I. Beloborod'ko and A.N. Omel'yanchuk, Sov. J. Low
Temp. Phys. {\bf 17}, 518 (1991).

\bibitem{Bel}  S.I. Beloborod'ko, private communicatrion.

\bibitem{Wolf}  E.L. Wolf, {\it Principles of Electron Tunneling
Spectroscopy}, Oxford University Press London 1985

\bibitem{Parks}  R.D. Parks, {\it Superconductivity}, Marcel Dekker Inc.
New York 1969.

\bibitem{Dolgov}  O.V. Dolgov, R.S. Gonnelli, G.A. Ummarino, A.A.
Golubov, S.V. Shulga, and J. Kortus,  Phys. Rev. B  {\bf 68},
132503 (2003). 

\bibitem{BTK}  G.E. Blonder, M. Tinkham and T.M. Klapwijk, Phys. Rev. B
{\bf 25}, 4515 (1982).

\bibitem{Naidyuk1}  Yu.G. Naidyuk, I.K. Yanson, L.V. Tyutrina, N.L.
Bobrov, P.N. Chubov, W.N. Kang, Hyeong-Jin Kim, Eun-Mi Choi, and
Sung-Ik Lee, JETP Lett. {\bf 75}, 283 (2002).

\bibitem{Kuz'menko}  A.B. Kuz'menko, F.P. Mena, H.J.A. Molegraaf,
D. van der Marel, B. Gorshunov, M. Dressel, I.I. Mazin, J. Kortus,
O.V. Dolgov, T. Muranaka, J. Akimitsu, Solid State Communication
{\bf 121}, 479 (2002).

\bibitem{Mazin2}  I.I. Mazin, O.K. Andersen, O. Jepsen, A.A. Golubov,
O.V. Dolgov, J. Kortus, cond-mat/0212417.

\bibitem{Szabo} P. Szab\'o ,P.\,Samuely, J.\,Kacmar\'cik, Th. Klein,
J. Marcus, D. Furchart, S. Miragila, C. Marcenat, A.G.M. Jansen,
Phys. Rev. Lett. {\bf 87} 137005 (2001).

\bibitem{Gonnelli} R.S. Gonnelli, D. Daghero, G.A. Ummarino,
V.A. Stepanov, J. Jun, S.M. Kazakov and J. Karpinski,
Phys. Rev. Lett. {\bf 89} 247004 (2002). 

\bibitem{Samuely} P. Samuely, P. Szabo, J. Kacmarcik, T. Klein,
A.G.M. Jansen, Physica C {\bf 385} 244 (2003).

\bibitem{Golubov2} A.A. Golubov, A. Brinkman O.V. Dolgov, J. Kortus,
and O. Jepsen, Phys. Rev. B {\bf 66}, 054524 (2002).

\bibitem{Naidyuk2}  Yu.G. Naidyuk, O.E. Kvitnitskaya, I.K. Yanson, S.-L.
Drechsler, G. Behr, and S. Otani, Phys. Rev. B {\bf 66}, 140301
(2002).

\bibitem{Heid} R. Heid, K.-P. Bohnen, and B. Renker, Adv. in Solid State
Phys. {\bf 42}, 293 (2002); R. Heid, B. Renker, H. Schober, P.
Adelmann, D. Ernst, and K.-P. Bohnen, Phys. Rev. B {\bf 67},
180510 (2003). 

\bibitem{Aizawa}  Takashi Aizawa, Wataru Hayami, and Shigeki Otani, Phys.
Rev. B {\bf 65}, 024303 (2001).

\bibitem{Bobrov} N.L. Bobrov, P.N. Chubov, Yu.G. Naidyuk, L.V. Tyutrina,
I.K. Yanson, W.N. Kang, Hyeong-Jin Kim, Eun-Mi Choi, C.U. Jung,
and Sung-Ik Lee, in book {\it  New Trends in Superconductivty},
Vol.67 of NATO Science Series II: Math. Phys. and Chem., ed. by J.
F. Annett and S. Kruchinin, (Kluwer Acad. Publ., 2002), p.225,
(see also: cond-mat/0110006).

\bibitem{Golubov} A.A. Golubov, J. Kortus, O.V. Dolgov, O. Jepsen, Y. Kong,
O.K. Andersen, B.J. Gibson, K. Ahn, and R.K. Kremer, J. Phys.:
Condens. Matter {\bf 14}, 1353 (2002).

\bibitem{Naidyuk3}  Yu.G. Naidyuk, I.K. Yanson, O.E. Kvitnitskaya, S.
Lee, and S. Tajima, Phys. Rev. Lett., {\bf 90}, 197001 (2003).

\bibitem{Kulik_therm}  I.O. Kulik, Sov. J. Low Temp. Phys. {\bf 18},
302 (1992).

\bibitem{Koshelev} A. E. Koshelev and A. A. Golubov,
Phys. Rev. Lett. {\bf 90}, 177002 (2003).
\end{thebibliography}
\end{document}